\documentclass[rsi, amsmath,amssymb,reprint, onecolumn]{revtex4-1}
\usepackage{graphicx}% Include figure files
\usepackage{dcolumn}% Align table columns on decimal point
\usepackage{bm}% bold math
\usepackage[utf8]{inputenc}
\usepackage[T1]{fontenc}
\usepackage{mathptmx}
\usepackage{etoolbox}

\usepackage{siunitx}
\usepackage{subcaption}
\usepackage{booktabs}
\usepackage{multirow}
\usepackage{physics}

\makeatletter
\def\@email#1#2{%
 \endgroup
 \patchcmd{\titleblock@produce}
  {\frontmatter@RRAPformat}
  {\frontmatter@RRAPformat{\produce@RRAP{*#1\href{mailto:#2}{#2}}}\frontmatter@RRAPformat}
  {}{}
}%
\makeatother
\begin{document}

\preprint{AIP/123-QED}

\title[]{The Diocotron Instability in the Trapped Electrons Experiment T-REX\\ and its Relevance to Electron Clouds in Gyrotron Guns}
% Force line breaks with \\
\author{F.~Romano*}
\email{francesco.romano@epfl.ch}
\author{P.~Giroud-Garampon}

\author{J.~Loizu}
\author{G.~Scimone}
%\author{S.~Antonioni}
%\author{M.~Noël}
%\author{J.~Genoud}
\author{J.-P.~Hogge}

%\author{T.~Goodman}
%\author{M.~Podestà}
%\author{F. Braunmüller}
\affiliation{\'Ecole Polytechnique Fédérale de Lausanne (EPFL), Swiss Plasma Center (SPC), CH-1015 Lausanne, Switzerland.}

\date{\today}% It is always \today, today,

\begin{abstract}

Gyrotrons are essential for electron cyclotron resonance heating (ECRH) in fusion reactors, making their efficient operation crucial for advancing fusion energy. Past experiments revealed instability issues due to trapped electrons in the magnetron injection gun (MIG) region, causing undesired currents and operational failures. To address this, tight manufacturing tolerances are required for the MIG geometry~\cite{pago2}. We present findings of the TRapped Electrons eXperiment (T-REX) developed at the Swiss Plasma Center, designed to understand the physics of electron clouds in gyrotron MIGs~\cite{Romano2024}. T-REX replicates MIG geometries, as well as their typical electric and magnetic fields, and it is supported by 3D Particle-in-Cell (PIC) simulations with the FENNECS code~\cite{guilPoP,guilPoP2,guilth,FENNECS,Pierrick}. The setup includes two coaxial electrodes in a vacuum chamber atop a superconducting magnet, with a central electrode biased to negative DC voltages and an outer one at ground, creating a radial electric field up to $\SI{2}{\mega\volt\per{\meter}}$ and an axial magnetic field $B < \SI{0.31}{\tesla}$. This setup mimics also the principle of Penning-Malmberg traps. Initial discrepancies between experiments and simulations were found to be directly linked to the diocotron instability and led to FENNECS being upgraded to 3D and a set of diagnostics for T-REX to be specifically designed. The diocotron instability leads the electron cloud to collapse and reform at a certain frequency that depends on the plasma conditions. Within this article, time-resolved current measurements on the main experiment components, namely the outer electrode and the top flange, are presented and discussed. Further on, a fast current probe array installed at the top flange is presented in detail. The measurements highlight rotating structures in the electron cloud resulting from the diocotron instability. Simulations show remarkable agreement with experiments, especially in terms of the frequency of build-up and collapse of the electron cloud, the rotation frequency and direction of the modes resulting from the diocotron instability. These results further improve our understanding of non-neutral plasmas in environments mimicking those of a real gyrotron MIG, paving the way for further improving gyrotron performance and reliability in fusion energy systems.

\end{abstract}

\maketitle

\section{\label{sec:introduction}Introduction}
\begin{figure}[h]
    \centering
    \includegraphics[width=.5\linewidth]{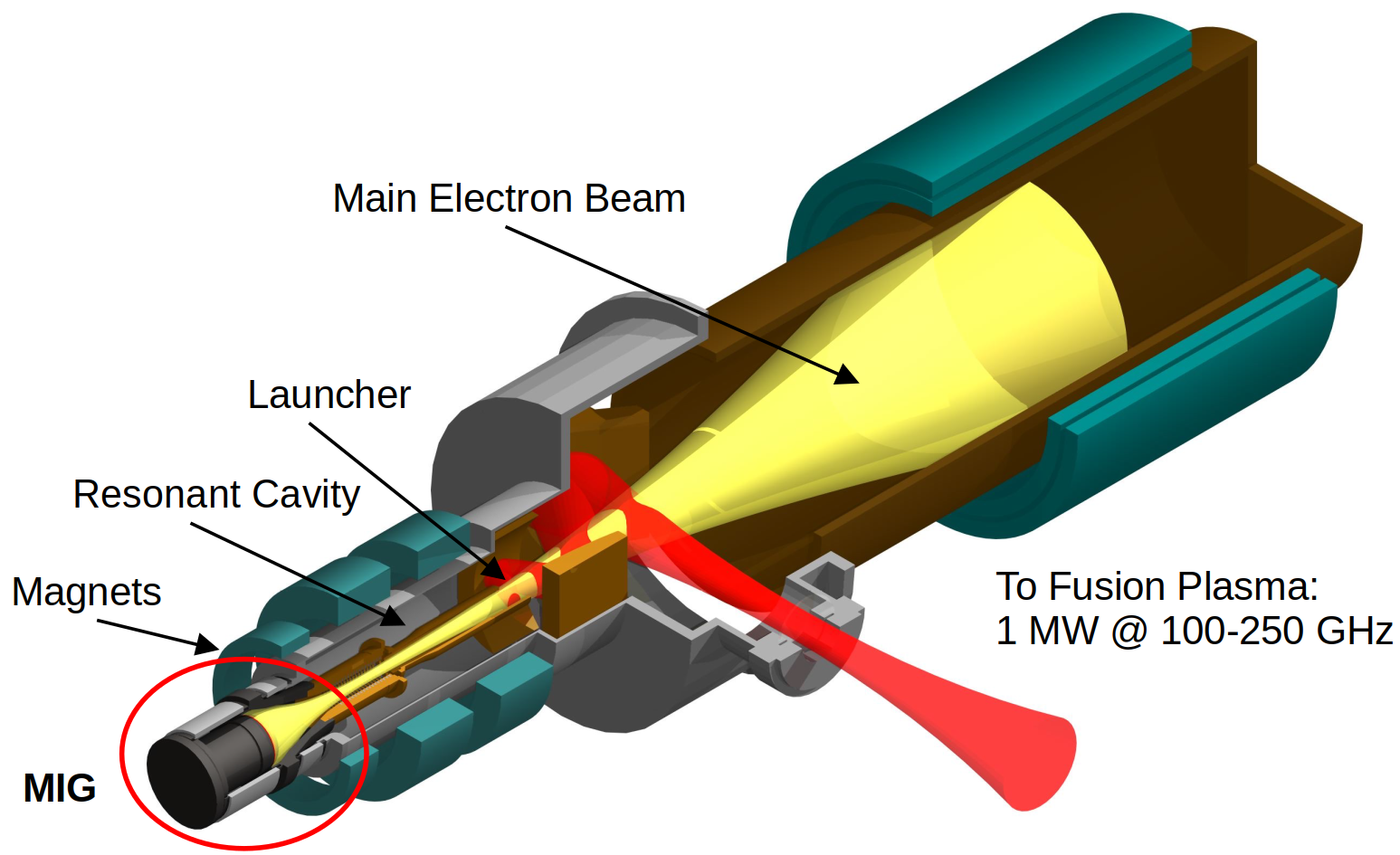}
   \caption{Schematics of a gyrotron and its main components. The red ellipse represents the gyrotron region where the secondary electrons (i.e. not belonging to the main electron beam) can potentially form a trapped electron cloud which is the object of this article. Adapted with permission from~\cite{alberti}: S. Alberti, J. Genoud, T. Goodman, J.-P. Hogge, L. Porte, M. Silva, T.-M. Tran, M.-Q. Tran, K. Avramidis, I. Pagonakis, et al., in EPJ
Web Conf., Vol. 157 (2017) p. 03001.}
    \label{fig:gyrotron}
\end{figure}

% % % %
To generate fusion energy, both tokamak and stellarator configurations foresee the use of gyrotrons as the main plasma heating devices via electron cyclotron resonance heating (ECRH) and/or current drive (ECCD)~\cite{stellarator,ITER}. The ITER fusion reactor currently requires 80 \SI{}{\mega\watt}-level gyrotrons~\cite{10157606,10157774}, operating at \SI{170}{\giga\hertz}. For DEMO, the current plan is to deliver \SI{130}{\mega\watt} of power via gyrotrons~\cite{DEMO}. Therefore, their efficient and reliable operation is critical, and any design modification that can ease construction requirements and tolerances will reduce costs and manufacturing time~\cite{pago1}. Gyrotrons are high-power devices that produce electromagnetic waves at frequencies in the sub-\SI{}{\tera\hertz} range. In a gyrotron, see Fig.~\ref{fig:gyrotron}, electrons are generated via thermionic emission from an emitter ring within the magnetron injection gun (MIG). The electrons are accelerated at mildly relativistic energies in a crossed configuration of electric and magnetic fields. They rotate at the relativistic cyclotron frequency, which depends mainly on the applied magnetic field amplitude, but also weakly on their energy. This defines the gyrotron operating frequency. The annular electron beam travels inside a cylindrical resonant cavity in which a specific electromagnetic wave mode is excited. The resonant interaction between the gyrating electrons and the electromagnetic mode supported by the cavity leads to a net transfer of energy from the electrons to the wave. After exiting the cavity, the excited transverse electric (TE) mode is converted into a Gaussian beam, which is then propagated through a diamond output window. Gyrotrons operate in ultra-high vacuum conditions $\sim\SI{1e-9}{\hecto\pascal}$.

This article focuses on the novel and unique non-neutral plasma experiment T-REX, that aims at understanding related issues found in some gyrotron MIGs. For specific conditions, nominal gyrotron operation can not be achieved and/or unexpected shutdown of the gyrotron can occur~\cite{pago2}. Specifically, unstable voltage standoff, arcs, electrode damage and localized melting have been observed. The cause has been tracked down to secondary electrons (i.e. electrons that do not belong to the main beam) trapped in the MIG region due to the presence of potential wells. Those can form when magnetic field lines cross twice an equipotential line. The potential well size and depth depend on the combination of the applied magnetic field and the geometry of the electrodes to which a potential is applied. These potential wells can trap electrons and, via electron leakage, can cause relatively large currents to arise, leading to possible internal gyrotron damage and preventing the power supply from sustaining the nominal, externally applied, voltage bias. The resulting currents can exceed the limits of the Body Power Supply (BPS), which is typically rated for \SI{40}{\kilo\volt} and \SI{150}{\milli\ampere}~\cite{gyroBPS}. While optimizing electrode geometry relative to the vacuum magnetic field can shape equipotential lines and suppress potential wells, such modification presents significant challenges in design complexity, fabrication costs, and development lean times. Furthermore, the limited diagnostic access in sealed gyrotrons requires dedicated plasma experiments to study trapped electron cloud dynamics in a controlled setting. Insights gained from such studies are essential for informing the design and fabrication of next-generation gyrotrons.

The remainder of this article is organized as follows. Section~\ref{sec:diocotron} examines non-neutral plasma theory and the diocotron instability within Magnetron Injection Gun (MIG) environments, detailing the mechanisms of plasma formation and sustenance and their subsequent impact on gyrotron performance. In Section~\ref{sec:TREX} and~\ref{sec:FENNECS} the primary research tools are presented, specifically the FENNECS simulation code and the T-REX experimental setup. Section~\ref{sec:diagnostics} provides a comprehensive description of the T-REX diagnostic suite, followed by a presentation of experimental results and a comparative validation against FENNECS simulations in Section~\ref{sec:results}. Finally, Section~\ref{sec:conclusions} draws conclusions and outlines future research directions.

\section{\label{sec:diocotron}Non-neutral Plasma, the Diocotron Instability, and its relevance to gyrotron's MIG}

\begin{figure}
    \centering
    \includegraphics[width=.3\linewidth]{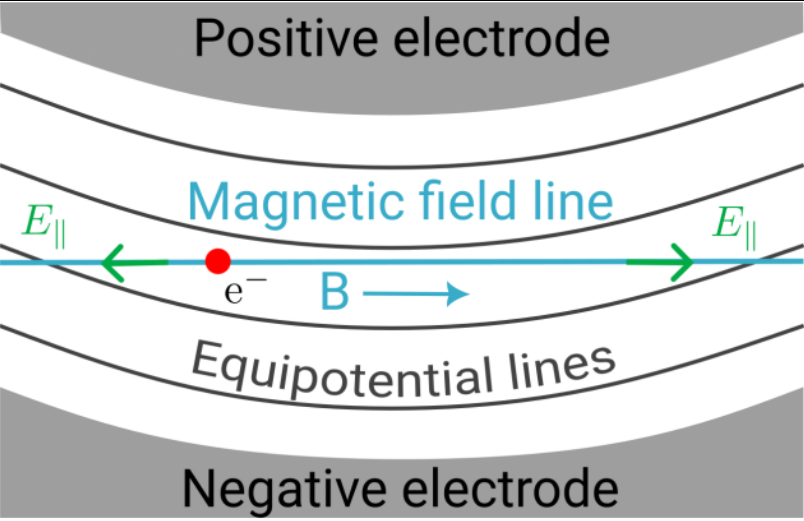}
\caption{Electron trapping via electric and magnetic fields: a magnetic field line is shown crossing twice an equipotential field line with an electron trapped. Adapted with permission from G. M. Le Bars, Modelling of non-neutral plasmas trapped by electric and magnetic fields relevant to gyrotron electron guns, Ph.D. thesis,
EPFL, Lausanne (2023)~\cite{guilth}.}
    \label{fig:trapping}
\end{figure}

The study of electron clouds and electron trapping belongs to the field of non-neutral plasmas systems, characterized by a single dominant species where global charge neutrality does not hold. This field has been studied extensively since the 1960s~\cite{Davidson1991PhysicsON}. Electrons can be trapped primarily through magnetic mirrors or a specific combination of electric and magnetic fields, namely those in which a magnetic field line intersects equipotential surfaces in such way that a potential well is formed along the field line, see Fig.~\ref{fig:trapping}. Due to the electric and magnetic field topology of a gyrotron's MIG, the resulting trapping mechanism belongs to the second case.

\begin{figure}
    \centering
    \includegraphics[width=.6\linewidth]{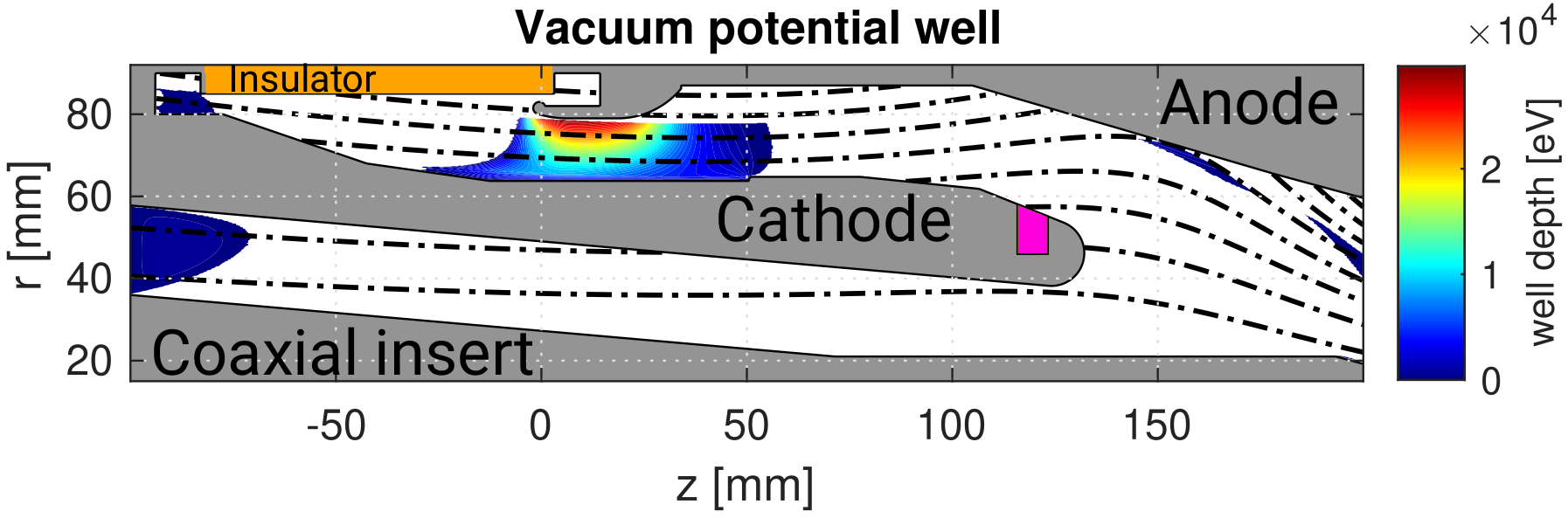}
\caption{Vacuum potential well in the MIG of the first European \SI{170}{\giga\hertz} - \SI{2}{\mega\watt} coaxial gyrotron prototype developed for ITER~\cite{pago3}. The cathode is at \SI{-60}{\kilo\volt} and the other electrodes (anode and coaxial insert) are at \SI{+30}{\kilo\volt}. The black dashed-dotted lines are the magnetic field lines, and the magenta surface highlights the emitter ring, where the main electron beam is produced. The $z-$axis (at $r=0$) corresponds to the axis of azimuthal symmetry. The colored areas in the vacuum region indicate different potential wells with the main one reaching a depth of \SI{30}{\kilo\electronvolt} as indicated by the color scale. Adapted with permission from G. M. Le Bars, Modeling of non-neutral plasmas trapped by electric and magnetic fields relevant to gyrotron electron guns, Ph.D. thesis,
EPFL, Lausanne (2023)~\cite{guilth}}
    \label{fig:MIG}
\end{figure}

A gyrotron's MIG, see Fig.~\ref{fig:MIG}, is characterized by a nearly axial magnetic field, determined by external magnets, and an electric field shaped by the electrodes biased at different voltages. In particular, the magnetic field is azimuthally symmetric and, while it has to be uniform in the resonant cavity of a gyrotron, it is non-uniform in the MIG region. The electric field, on the other hand, is mainly radial and imposed by the different electrodes. In the MIG region, vacuum potential wells arise where the magnetic field lines cross twice the equipotential lines, see Fig.~\ref{fig:trapping}~\cite{pago2}.

In MIGs, the trapped electrons do not belong to the main gyrotron electron beam, but are created by ionization of the residual gas. Specifically, the thermoionically generated - primary - electrons are accelerated from the emitter ring towards the gyrotron's resonant cavity. The crossed electric and magnetic fields allow to impart both perpendicular (to the magnetic field) and parallel kinetic energy. At nominal conditions, primary electrons (in yellow on Fig.~\ref{fig:gyrotron}) leave the emitter and eventually reach the gyrotron collector. The secondary electrons generation mechanisms, instead, can be diverse:
\begin{enumerate}
    \item Cosmic rays collide with neutrals and generate free electrons;
    \item Field emission: electrons emitted from the biased surfaces due to large electric fields;
    \item Ion induced electron emission (IIEE): free electrons can be emitted from ion impact on the electrode surfaces; 
    \item Electron impact ionization of neutrals: free electrons impact neutrals releasing free electrons.
\end{enumerate}

Once a first electron is created in the potential well, it will rapidly reach the azimuthal drift velocity due to the $\vec{E}\times\vec{B}$ that is in the range of $\SIrange{1E6}{1E7}{\meter\per\second}$. At these kinetic energies, electrons can ionize residual gas particles via impact ionization, generating more electrons in the potential well that will sustain the electron plasma. As the amount of charge increases in the potential well, the space charge of the electron cloud also modifies the potential well, and, as a result, it can reach electron densities that can create a path for undesired currents to flow via the electron cloud between the electrodes surfaces in a gyrotron's MIG. Such leakage currents cause problems and damages, in particular unstable voltage standoff, arcs, and even localized melting~\cite{1321243}. Modern design criteria for gyrotron MIGs mandate the elimination of vacuum potential wells by tailoring electrode geometries to strictly adhere to the nominal magnetic field topology. Satisfying these constraints imposes substantial engineering and manufacturing overhead, typically necessitating a computationally intensive, iterative optimization process. Moreover, this rigid alignment inherently restricts the gyrotron's operational envelope by "freezing" the permissible magnetic field configurations, thereby limiting the device's flexibility across different operating regimes.

In these configurations of crossed electromagnetic fields and non-neutral plasmas, the dynamics of such electron clouds are complex, in particular due to the diocotron instability, which is driven by the radial shear in the azimuthal $\vec{E} \times \vec{B}$ flow~\cite{guilth, Eggleston, Danielson}.

Only recent investigations strongly suggested that indeed this instability may be the primary mechanism causing bursty or disruptive behaviors in coaxial cavities, specifically within gyrotrons MIGs~\cite{Pierrick}. 

This instability represents the plasma-physics analogue of the Kelvin-Helmholtz instability in classical fluid dynamics, where the transverse gradient in the velocity field provides the free energy for wave growth. The diocotron instability is characterized by an azimuthal structure that often decomposes into a few dominant modes labeled by their mode number $m$. In the linear regime, radial density gradients within the plasma column generate self-consistent electric fields that amplify initial perturbations, leading to the exponential growth of propagating azimuthal waves. As the instability enters the non-linear regime, it induces significant cross-field transport, leading to a redistribution of the equilibrium density. This process often results in the "smearing" of the initial shear layer or the formation of stable, coherent vortex structures. Such dynamics can eventually lead to the saturation of the instability or even reorganize the plasma into a more stable, quiescent state. In the context of a gyrotron MIG, these non-linear effects are critical, as they dictate the ultimate density of the trapped electron cloud and, consequently, the magnitude of the resulting parasitic leakage currents.

Future gyrotrons will need to operate at significantly higher power levels, which will certainly require coaxial designs such as the one shown in Fig.~\ref{fig:MIG}. In these configurations, the problem of trapped electron clouds becomes considerably more severe. These observations, together with the distinctive and largely unexplored parameter regimes characteristic of such MIGs, provide strong motivation for a dedicated scientific investigation of the underlying physics of trapped electrons. A deep understanding of these phenomena is essential to enable improved MIG designs and to develop targeted mitigation strategies for next-generation high-power gyrotrons.

\section{\label{sec:TREX}The TRapped Electrons eXperiment - T-REX}

\begin{figure}[h]
    \centering
    \includegraphics[width=.8\linewidth]{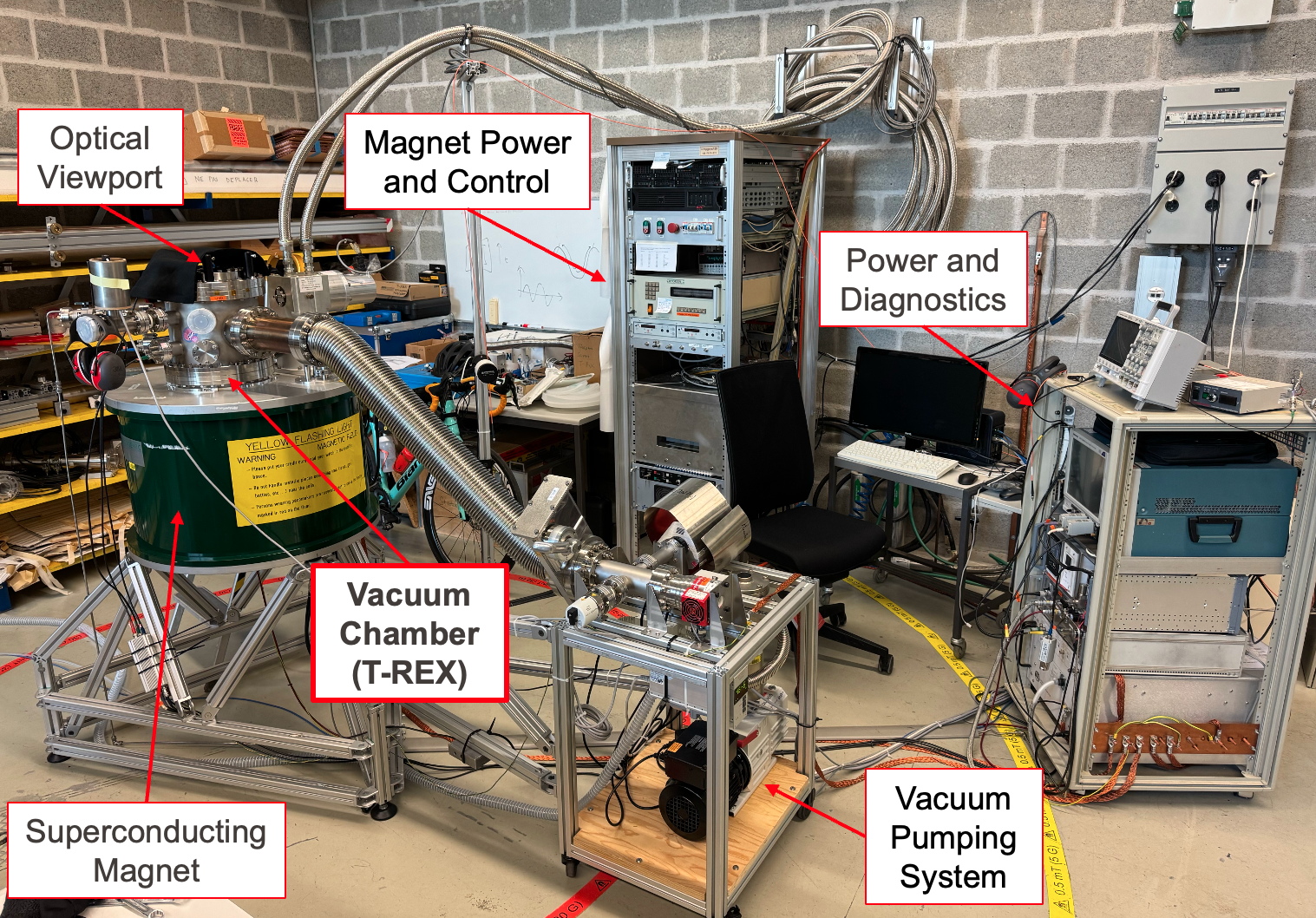}
   \caption{Picture of the T-REX facility. Left: superconducting magnet with the vacuum chamber containing T-REX on top. Behind: the magnet power and controls rack. Middle front: pumping system. Right: power and diagnostics rack.}
    \label{fig:T-REX_Photo}
\end{figure}

T-REX is a unique basic plasma experiment aiming at studying electron cloud formation in gyrotron's MIG-like configurations without the presence of primary electrons. The concept is to recreate environments similar to those that are problematic in a gyrotron's MIG, namely a potential well produced by specific topologies of electric and magnetic fields, electrodes geometries, background pressures, and gas compositions. An image of the T-REX setup is shown in Fig.~\ref{fig:T-REX_Photo}.

\begin{figure}[h]
    \centering
    \includegraphics[width=.7\linewidth]{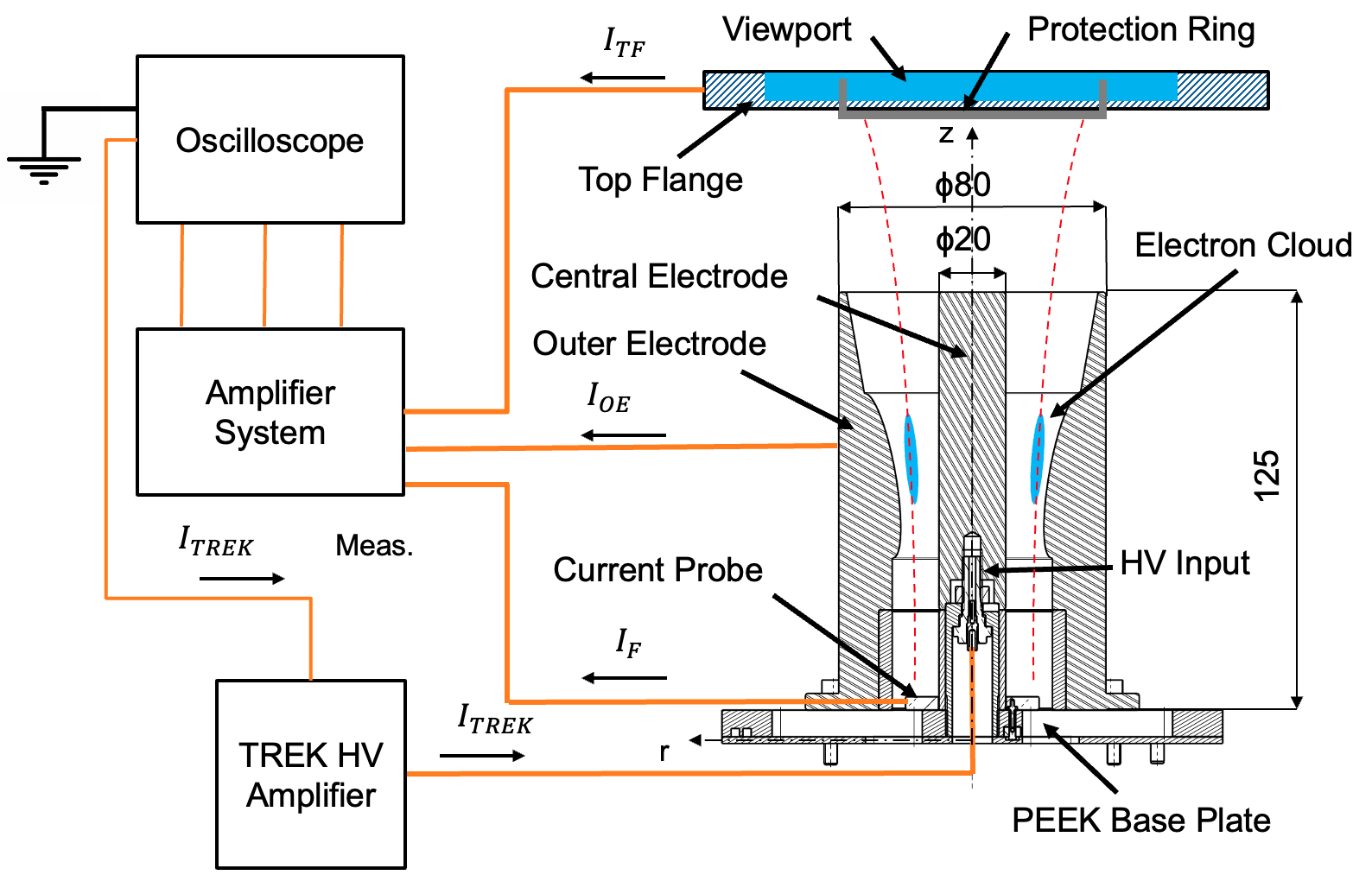}
    \caption{Sectioned view of T-REX electrodes and current measurement set-up. Dimensions are in [\SI{}{\milli\meter}]. The figure highlights the amplifier system that measures the current collected at different locations in the experiment: at the top flange $I_\text{TF}$, at the outer electrode $I_\text{OE}$, and at the current probe $I_\text{F}$. Those values together with the total input current $I_\text{TREK}$ are acquired via an oscilloscope.} % definition of electron cloud geometry, $w$ radial thickness, $h$ vertical height. 
    \label{fig:measure}
\end{figure}

The core of the experiment, see Fig.~\ref{fig:measure}, is composed of two coaxial electrodes installed within a vacuum chamber placed on top of a superconducting magnet. The electrodes are shaped such as to form a potential well mimicking those existing in gyrotrons MIGs, with the central electrode biased at a negative high-voltage (HV) and the outer electrode at ground. The electrodes geometries define the vacuum electric field, while the magnetic field shape, slightly divergent in the trapping region, is determined by the superconducting magnet. The amplitude of the magnetic field in the area of the electron cloud reaches a maximum of $B<\SI{0.4}{\tesla}$. To study the electron cloud behavior on multiple environments, gas injection is available. This can also be used to regulate the pressure $p$ inside the vacuum chamber in the range $\SI{1E-6}{} \le p \le \SI{5E-4}{mbar}$. The HV is delivered to T-REX via a TREK HV amplifier 20/20C that provides up to $\pm \SI{20}{\kilo\volt}$ DC at up to $\pm \SI{20}{\milli\ampere}$. The coaxial electrodes of T-REX are made of 6082 aluminum alloy and are interchangeable, allowing for multiple geometries to be tested. At the base of the electrodes, an aluminum ring is installed which operates as a current probe to measure the amount of electrons that escape downwards, see Fig.~\ref{fig:measure}. The electrodes assembly is mounted on top of a polyether ether ketone (PEEK) plate which electrically isolates it from the vacuum chamber, and also provides guiding for the wiring to the electrical feedthroughs of the vacuum chamber. The design of T-REX has been supported by simulations performed with FENNECS~\cite{guilPoP,guilPoP2,guilth,FENNECS,Pierrick}. 

\begin{figure}[h]
    \centering
    \includegraphics[width=.85\linewidth]{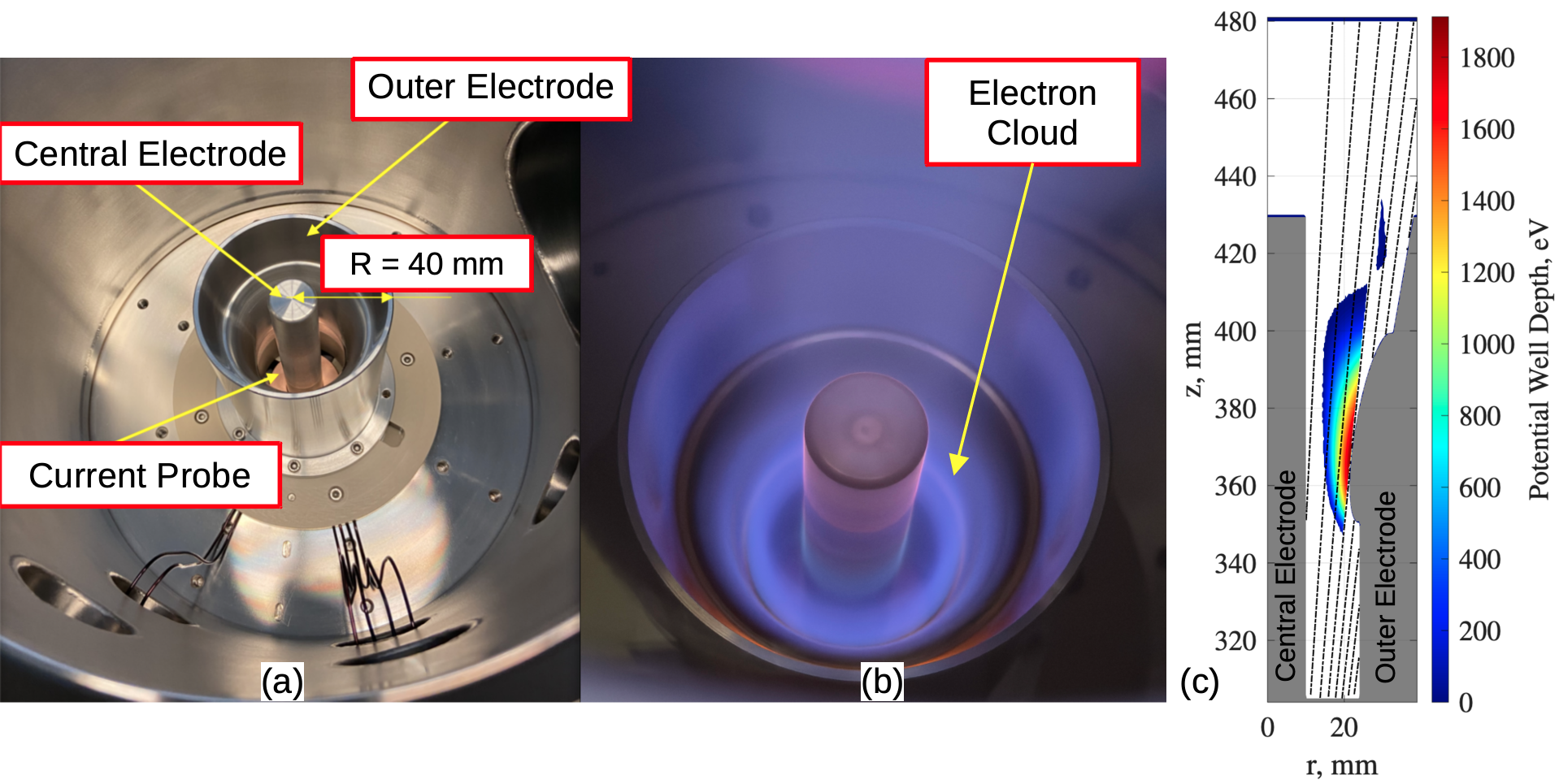}
    \caption{Inside view of T-REX: \textit{(a)} two coaxial electrodes, current probe, PEEK insulating base plate and cabling. \textit{(b)} T-REX in operation $B=\SI{0.31}{\tesla}$, $\Delta V_\text{bias}=\SI{10}{\kilo\volt}$, He at $p=\SI{3E-4}{\hecto\pascal}$ highlighting the glowing electron cloud. \textit{(c)} Vacuum potential well within T-REX for $B=\SI{0.10}{\tesla}$, $\Delta V_\text{bias}=\SI{10}{\kilo\volt}$, central and outer electrode in grey, dashed lines represent the magnetic field lines, color map shows the potential well depth in \SI{}{\electronvolt}. }
    \label{fig:T-REXin}
\end{figure}

At the top of the experiment, a DN160 borosilicate viewport with an indium tin oxide (ITO) coating is installed. The coating provides electrical conductivity to help evacuate electrons that impact on it. In front of the viewport, on the vacuum side, a stainless steel protection ring is installed to shield the viewport region that receives most of the high intensity electron flow during operation. Significant damage to the viewports has previously been observed in the absence of shielding. In Figure~\ref{fig:T-REXin}(a) the T-REX electrode assembly is shown highlighting the central and outer electrode, the current probe at the bottom, and the cabling connecting to the electrical vacuum feedthroughs for measurement purposes. In Figure~\ref{fig:T-REXin}(b), T-REX is depicted during operation: a ring-shaped electron cloud spontaneously forms and is visible near the outer electrode, as simulations predict. The plasma light emission from the electron cloud is the result of mostly excitation and de-excitation of helium (in this case). In a precedent work~\cite{Romano2024}, we did an estimate based on cross-sections for different processes (ionization, recombination, excitation) and the very rapid loss of ions to the central electrode rules out recombination as a significant contributor to the light emission. A distinct luminous layer is also observed at the central electrode interface, likely resulting from intense thermal loads and charged-particle bombardment during long high-current operation. Preliminary optical emission spectroscopy (OES) measurements of this localized emission were inconclusive.

Previous simulations of T-REX~\cite{guilPoP,guilPoP2,guilth,FENNECS,Pierrick} showed that, with the available combination of magnetic and electric field topology (electrodes geometry), background gas pressures and compositions, the resulting potential well forming between the two electrodes traps electrons and leads to an electron cloud with the following properties:
\begin{itemize}
    \item Electron kinetic energy $E_k = \SI{1}{\electronvolt}-\SI{1}{\kilo\electronvolt}$, $v_{\vec{E} \times \vec{B}, max}=\SI{2E7}{\meter\per\second}$
    \item Maximum electron cloud density range: $n_{e,max}=\SI{1E15}{}-\SI{2E16}{\meter^{-3}}$%, 
   % \item Electron cloud (radial) thickness: $w=\SI{1.5}{\milli\meter}$, vertical height: $h=\SI{45}{\milli\meter}$, see Fig.~\ref{fig:measure}.%, Larmor radius $\rho_L \sim \SI{0.05}{\milli\meter}$
    \item Brillouin ratio $2\omega_p^2/\omega_c^2 = 0.1-0.2$ 
\end{itemize}

where $\omega_p = \sqrt{{n_e q_e^2} / {m_e \epsilon_0}}$ is the plasma frequency, $v_{\vec{E} \times \vec{B}, max}$ the maximum drift velocity due to the $\vec{E}\times\vec{B}$ drift, $\omega_c= q_e B / m_e$ is the electron cyclotron frequency, with $n_e, m_e, q_e$ the electron density, mass, and charge respectively, and $\epsilon_0$ the vacuum permittivity. The Brillouin ratio measures the relative strengths of the (de-focusing) space-charge force and the (focusing) magnetic force on the plasma~\cite{Davidson1991PhysicsON}, and in this case indicates that the electron clouds obtained in T-REX have large densities compared to Penning traps that usually result in Brillouin ratios much smaller than $1$~\cite{guilth}.

For the first set of tests, an amplifier system has been applied to obtain a time-resolved measurement of the top flange $I_\text{TF}$ and outer electrode $I_\text{OE}$ currents. The complete top flange assembly is electrically isolated from the rest of the vacuum chamber, and is grounded via an amplifier system to measure $I_\text{TF}$, hence we get an integrated measurement of the current collected on the top of T-REX. The outer electrode is also electrically isolated and grounded via the same amplifier system to measure $I_\text{OE}$, the current collected on the outer electrode, see Fig.~\ref{fig:measure}. The system is driven by a shunt-amplifier system that has a cutoff frequency of $f_c \sim \SI{437}{\kilo\hertz}$ and an accuracy of $\pm\SI{0.05}{\milli\ampere}$. The current flowing from the current probe - bottom of the electron assembly - to the ground is measured via an integrator circuit. Its amplitude is observed to be typically small $I_\text{F} < \SI{5}{\micro\ampere}$ compared to $I_\text{TF}$, $I_\text{OE}$ in the \SI{}{\milli\ampere} range and, even though always measured, is excluded from this analysis. The current delivered by the TREK HV amplifier, $I_\text{TREK}$ - the total input current - is measured via the HV amplifier itself with a time-resolution $>\SI{10}{\kilo\hertz}$, therefore only the average value is taken. The output voltage from the TREK, or the applied bias $\Delta V_\text{bias}$ is also measured via the HV amplifier itself with a time-resolution $>\SI{10}{\kilo\hertz}$. A voltage probe is also installed after the output of the HV amplifier within an in-house made protection circuit also including an HV switch. Both voltage measurements are taken as indicators to monitor the correct voltage output. Future upgrades shall include the installation of time-resolved voltage probe as close as possible to the electrodes to observe time-behaviours. Finally, the sum of currents and voltages are visualized and acquired via a \SI{33}{\giga\hertz} bandwidth oscilloscope. A second, \SI{350}{\mega\hertz} bandwidth oscilloscope, is used to monitor the applied bias $\Delta V_\text{bias}$. Finally, the experiment control parameters are:
\begin{itemize}
    \item Magnetic field $B\sim 0.1-\SI{0.4}{\tesla}$ ; 
    \item Electric field via the applied voltage to the central electrode $\Delta V_\text{bias}=0-\SI{20}{\kilo\volt}$;
    \item Background pressure $p\sim \SI{1E-6}{}-\SI{1E-3}{\hecto\pascal}$;
    \item Gas composition Ar, H$_2$, He, N$_2$, Ne;
    \item Electrodes geometry.
\end{itemize}

\subsection{Formation and dynamics of the cloud and the diocotron instability in T-REX}

This section provides further details on the physics of formation and dynamics of electron clouds in T-REX. 

In T-REX, a single free electron entering the potential well, originating from sources such as cosmic rays, field emission, or a finite ionization rate at room temperature, as explained in Sec.~\ref{sec:diocotron}, can trigger the electron density growth. This initial electron quickly acquires the $\vec{E} \times \vec{B}$ drift velocity in the azimuthal direction, corresponding to a kinetic energy of order \SI{100}{\electronvolt} under typical T-REX conditions, and starts ionizing the background neutral gas via impact ionization. At these energies, the electron-impact ionization cross-section is of order \SI{1e-20}{\meter^{-2}} for argon, giving an ionization frequency of order \SI{1e5}{\second^{-1}} at the typical operating pressure in T-REX. During impact ionization, the available kinetic energy is shared between the scattered and newly created electrons, with the latter typically born at low energy (of order 1--\SI{10}{\electronvolt}). The resulting electrons are therefore immediately trapped in the potential well, they reach the $\vec{E} \times \vec{B}$ drift velocity and further contribute to the ionization process. 

The ions, instead, due to their large Larmor radius and the large negative potential at the central electrode, are rapidly lost upon impacting it, on a timescale below \SI{100}{\nano\second} according to FENNECS simulations. Their impact releases additional electrons via ion-induced electron emission (IIEE). These electrons are born outside the potential well region of T-REX, see Fig.~\ref{fig:T-REXin}, and escape axially toward the top flange (TF) of the experiment where they are collected. In such electrodes configuration, recombination is negligible due to the short ion life $<\SI{100}{\nano\second}$. This process leads to the formation of a high-density electron cloud in the potential well. Collisions between cloud electrons and neutral gas generate a net azimuthal drag force acting on the electrons, which, in combination with the axial magnetic field, induces a radial drift of the electron cloud towards the outer electrode. This drift results in electron loss at the outer electrode, though some electrons may escape earlier if their axial kinetic energy is larger than the depth of the potential well.  According to 2D FENNECS simulations~\cite{Pierrick}, a steady-state can thus be achieved when the cloud density reaches a point where the impact ionization rate is balanced by axial and radial drift losses. Consequently, currents are expected on the outer electrode, due to collisional radial drift losses, and on the top flange due to axial losses from the cloud and from IIEE at the central electrode, where electrons produced by impacting ions are not trapped. However, according to 3D FENNECS simulations, the diocotron instability arises beyond a critical density and, before such steady-state is reached, causes the electron cloud to collapse. As the applied bias and magnetic field are supplied continuously, the electron cloud can self-reform. This phenomenon repeats cyclically, with a period typically on the order of hundreds of \SI{}{\kilo\hertz}. At the moment of collapse, the electron cloud loses confinement and a significant fraction of the total charge is expelled axially and radially, towards the top of the vacuum chamber, and to the outer electrode respectively. The diocotron instability evolves on timescales set by the characteristic $\vec{E}\times\vec{B}$ rotation frequency $f_{\vec{E} \times \vec{B}}$, which is of order $\sim \SIrange{10}{100}{\mega\hertz}$ under T-REX conditions. Its development leads to the formation of azimuthal mode structures - namely regions with larger electron plasma density - that rotate azimuthally with frequencies scaling with the $\vec{E}\times\vec{B}$ drift, typically as $f \sim m \cdot f_{\vec{E} \times \vec{B}}$, reaching values up to hundreds of~\SI{}{\mega\hertz} depending on the mode number $m$. 

\subsection{\label{sec:diagnostics}T-REX Current Probe Array System}

\begin{figure}[h]
    \centering
    \includegraphics[width=.8\linewidth]{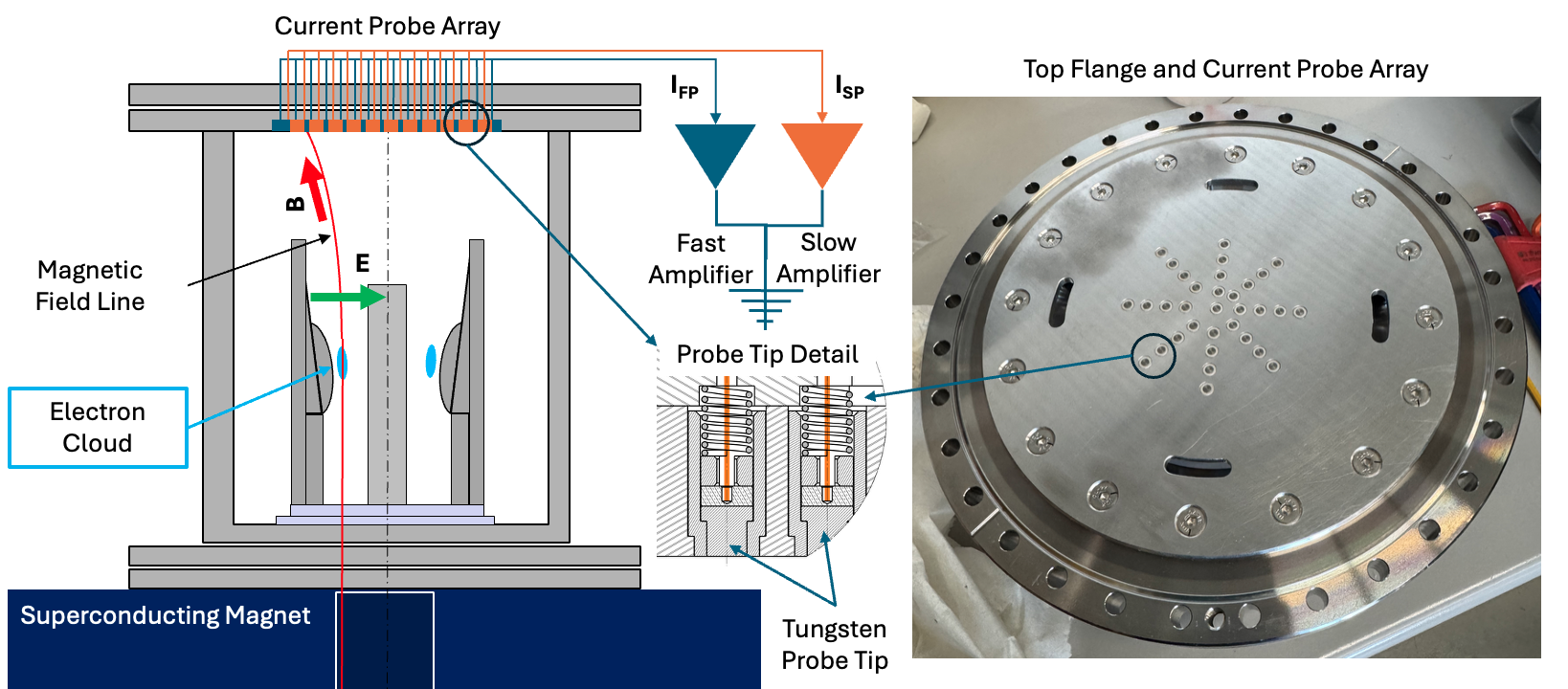}
    \caption{Current probe array schematics: the figure shows their installation in T-REX with a general schematics of the measurement circuit with "fast" probes $I_\text{FP}$ and "slow" probes $I_\text{SP}$. In the middle a detailed view of the probe's tip is displayed. On the right, a photo of the current probe array installed within the top flange is shown, the probe's disposition along a spiral trajectory can be evinced.}
    \label{fig:probe_array}
\end{figure}

This section describes the T-REX current probe array measurement system that has been employed for obtaining the higher spatial and temporal resolution results presented in this article. Indeed, to gain additional details on the physics of the electron clouds in T-REX more advanced diagnostics are required, which go beyond the initial measurement system installed. To measure the radial current distribution and detect signatures of the diocotron instability - which are expected to reveal themselves as azimuthal structures in the electron cloud - we developed a current probe array to be installed at the place of the existing viewport, see Fig.~\ref{fig:probe_array}. The array consists of 33 planar cylindrical probes made of tungsten, each of them with a diameter of \SI{6.5}{\milli\meter} facing the plasma. The probes follow a spiral distribution to ensure uniform radial coverage. Due to the expected small amount of charge collected on each probe, a dedicated amplification system is employed. In a first stage, 3 probes are equipped with a fast amplification system to measure the "fast current", $I_\text{FP}$, with a cutoff frequency of $f_c ~ > \SI{0.8}{\giga\hertz}$. The remaining probes are planned to operate with a cutoff of $f_c > \SI{400}{\kilo\hertz}$ returning a "slow current", $I_\text{SP}$. The number of fast probes will be increased in a second stage, see Fig.~\ref{fig:probe_array}. For the results presented in the article, only the fast probes were used, as the slow probe diagnostics were unavailable during the experimental campaign.

The installed amplifier system has been tested via a vector network analyzer (VNA) to determine its frequency response. To further validate the set-up, the frequency response has also been measured for the different circuit segments: from the probe to the vacuum electrical feedthrough, to the cable connecting to the oscilloscope. The 3 "fast" channels have a cutoff frequency of $f_c \sim \SI{1}{\giga\hertz}$ and maintain a gain of $\sim 24$ over the range of frequencies of interest. The amplifier system is measured to maintain the same $f_c$ for a minimum current less than  $I_{\text{FP,min}} < \SI{3.8}{\micro\ampere}$ and a maximum one of $I_\text{FP,max} < \SI{0.84}{\milli\ampere}$, based on the installed amplifier and an input resistance of \SI{100}{\ohm}. It should be noted that the primary objective of this analysis is the detection of mode rotation; consequently, temporal precision is prioritized, while amplitude variations—though accurately recorded—are not the subject of investigation in the scope of this article. 

\section{\label{sec:FENNECS}The FENNECS Code}

The simulations presented in this work are performed using the FENNECS code~\cite{FENNECS}, a kinetic electrostatic particle-in-cell (PIC) solver developed to model magnetized non-neutral plasmas confined by electrodes of complex geometry. Initially developed in axisymmetric geometry, the code has been extended to 3D to capture intrinsically non-axisymmetric phenomena such as the diocotron instability~\cite{letterPierrick,Giroud2026}.

FENNECS self-consistently solves the Boltzmann–Poisson equations for the electron distribution function $f_e(\vec{r},\vec{v},t)$ and the electrostatic potential $\phi(\vec{r},t)$,
\begin{equation} 
    \left[\pdv{t}+\vec{v}\cdot\pdv{\vec{r}}-\frac{q_e}{m_e}\left(\vec{E}+\vec{v}\times\vec{B}^{\text{ext}}_0(\vec{r})\right)\cdot\pdv{\vec{v}}\right]f_e(\vec{r},\vec{v},t) = C_{e,n}(f_e) + S_\text{seed} + S_\text{IIEE} + S_\text{SEE} - L_\text{wall}, \label{eq:boltzmann} 
\end{equation} 
\begin{equation} 
    \nabla^2\phi(\vec{r},t)=\frac{q_e}{\varepsilon_0}\int f_e(\vec{r},\vec{v},t) d^3\vec{v}, 
    \label{eq:poisson} 
\end{equation} 
under externally imposed magnetic fields. The plasma-generated magnetic field is neglected compared to the strong externally imposed magnetic field, an approximation that is well justified in the parameter regime relevant to T-REX. Electron dynamics are modeled using a PIC approach~\cite{Birdsall2018}, with particle trajectories advanced using the Boris algorithm~\cite{Boris1970}.

Here, $C_{e,n}$ denotes electron--neutral collisions with a uniform, cold background gas, that are modeled using a Monte Carlo approach~\cite{Birdsall1991}. Only elastic scattering and single-impact ionization are retained, while excitation processes are neglected due to their comparatively small cross-sections at the relevant electron energies. Ionization events generate secondary electrons, whereas the resulting ions are rapidly lost to the electrodes due to their large Larmor radius. The trajectories of the ions are nevertheless followed to account for ion-induced electron emission (IIEE) upon impact on metallic surfaces~\cite{Hasselkamp1992}, represented by the term $S_{\mathrm{IIEE}}$, which constitutes an important source of electrons in high-voltage configurations. Secondary electron emission due to electron impact on electrodes is also included through the source term $S_{\mathrm{SEE}}$, although it does not appear to play a significant role under T-REX operating conditions. A volumetric seed source $S_{\mathrm{seed}}$ can be introduced to initiate electron cloud formation, and particle losses at material boundaries are accounted for through the term $L_{\mathrm{wall}}$.

FENNECS is implemented in cylindrical coordinates $(r,\theta,z)$. To efficiently solve the Poisson equation, a spectral decomposition is employed in the azimuthal direction, combined with a finite-element discretization based on weighted extended B-splines in the $(r,z)$ plane~\cite{FENNECS}. This approach enables an accurate representation of complex axisymmetric geometries while retaining a structured grid~\cite{Hollig2001}.

The 2D version of FENNECS was previously validated against experiments and used to guide the design of T-REX~\cite{guilPoP,guilPoP2,Pierrick}. The 3D implementation has been verified by simulating the diocotron instability in simplified configurations, showing quantitative agreement with analytical linear theory~\cite{Davidson1991PhysicsON} and providing confidence in the physical and numerical accuracy of the model~\cite{Giroud2026}. The results presented in this article also serve as a validation of FENNECS 3D against experimental measurements. 

\section{\label{sec:results}T-REX Experimental Results and FENNECS 3D Validation}

We present the experimental conditions and results for T-REX, obtained using the electrode geometry illustrated in Fig.~\ref{fig:measure}. First, we employ the current distribution measurement system for $I_\text{OE}$ and $I_\text{TF}$ - with a cutoff frequency $f_c \sim \SI{437}{\kilo\hertz}$ - to characterize both the spatial partitioning of the current between the outer electrode and the top flange, as well as its temporal evolution. Second, we utilize the high-speed current probe array system to resolve fast-rotating structures within the plasma. 

The primary experimental parameters include the applied bias $\Delta V_\text{bias}$, magnetic field $B$, background pressure $p$, and gas species (Ar or H$_2$). Three magnetic field configurations were characterized, corresponding to $B\sim 0.10, 0.17, \SI{0.31}{\tesla}$ within the electron cloud region. A constant flow of either Ar or H$_2$ was maintained to establish the background pressure. It should be noted that the recorded pressure serves as an approximation of the vacuum level, as the measurement was not taken directly within the cloud region. Applied biases ranged from the minimum level to start the plasma ($\Delta V_\text{bias} \approx 1.5 - \SI{2}{\kilo\volt}$) to the maximum achievable applied bias $\Delta V_\text{bias}$, limited by the TREK HV amplifier’s current capacity of $I_\text{TREK,max} = \pm\SI{20}{\milli\ampere}$.

\begin{figure}[h]
    \centering
    \includegraphics[width=1\linewidth]{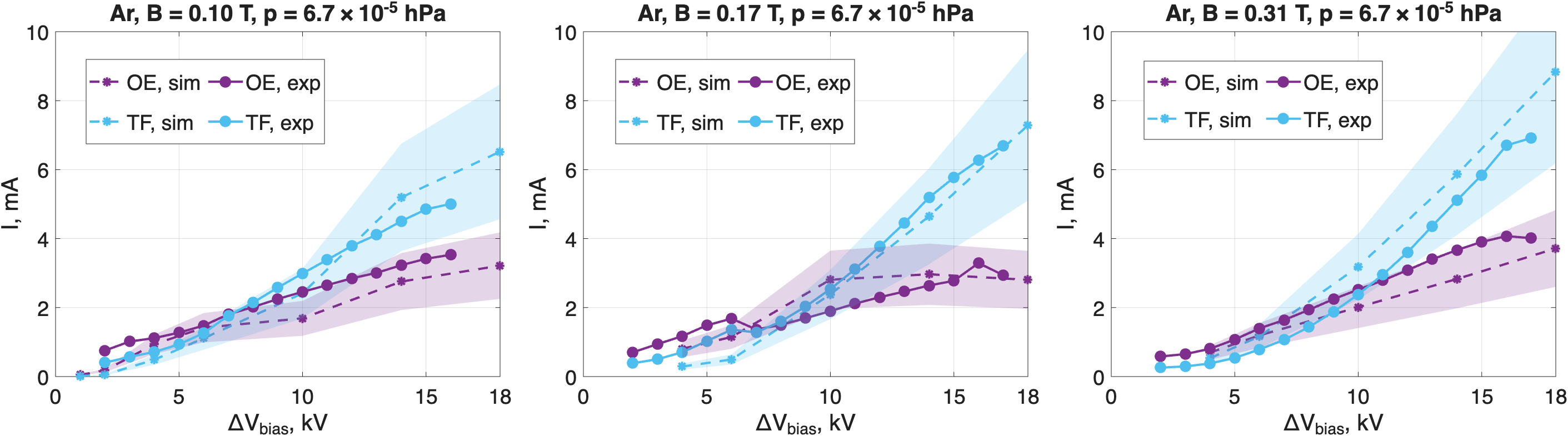}
    \caption{Comparison between T-REX experimental measurements (continuous lines) and FENNECS simulations (dashed lines) of outer electrode $I_\text{OE}$ (purple) and top flange $I_\text{TF}$ (light blue) currents vs applied bias $\Delta V_\text{bias}$ for three different magnetic fields $B$ and Ar as working gas. Color bands represent the uncertainty in the simulation due to rescaling simulation results with the experimental measured pressures that have an uncertainty of at least $\pm 30\%$.}
    \label{fig:current_spanAr}
\end{figure}

\begin{figure}
    \centering
    \includegraphics[width=1\linewidth]{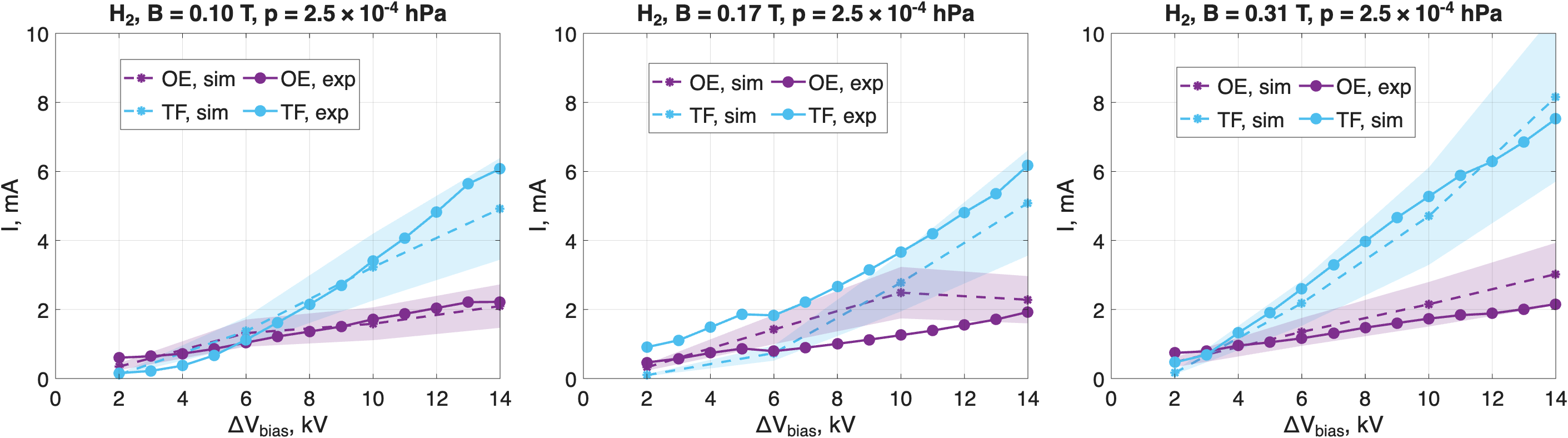}
    \caption{Comparison between T-REX experimental measurements (continuous lines) and FENNECS simulations (dashed lines) of outer electrode $I_\text{OE}$ (purple) and top flange $I_\text{TF}$ (light blue) currents vs applied bias $\Delta V_\text{bias}$ for three different magnetic fields $B$ and H$_2$ as working gas. Color bands represent the uncertainty in the simulation due to rescaling simulation results with the experimental measured pressures that have an uncertainty of at least $\pm 30\%$.}
    \label{fig:current_spanH2}
\end{figure}

We first characterize the T-REX operating regime in terms of time-averaged current amplitudes collected on the different parts of the device and their dependence on the applied bias $\Delta V_\text{bias}$, magnetic field amplitude $B$, pressure $p$, and gas. Fig.~\ref{fig:current_spanAr} (for Ar), and~\ref{fig:current_spanH2} (for H$_2$) show the time-averaged currents measured at the top flange, $I_\text{TF}$, and at the outer electrode, $I_\text{OE}$ for three different magnetic fields over a range of applied bias $\Delta V_\text{bias}$. 

The results highlight the increase in current amplitudes by increasing the applied bias $\Delta V_\text{bias}$. While the amplitude of $I_\text{TF}$ scales significantly with the magnetic field $B$, $I_\text{OE}$ exhibits a similar but markedly less pronounced trend. Furthermore, $I_\text{TF}$ tends to be larger than $I_\text{OE}$ with increased separation at larger applied bias $\Delta V_\text{bias}$. Currents are slightly larger in amplitude for Ar compared to H$_2$ due to the larger ionization cross-section. Those results are directly compared with FENNECS 3D simulation results, shown with dashed lines in Fig.~\ref{fig:current_spanAr} and Fig.~\ref{fig:current_spanH2}, and the excellent agreement confirms the ability of the code to accurately reproduce the observed T-REX behavior. The accuracy of the current amplitude measurement is of $\pm\SI{0.05}{\milli\ampere}$. The uncertainty (color bands) region in the simulation is due to the rescaling of the simulation results with the experimental pressure that has an uncertainty of at least $\pm 30\%$. It should be noted, however, that the pressure is measured at the side of the vacuum chamber, which surely differs from the pressure inside the electrodes cavity.

\begin{figure}[h]
    \centering
    \includegraphics[width=.85\linewidth]{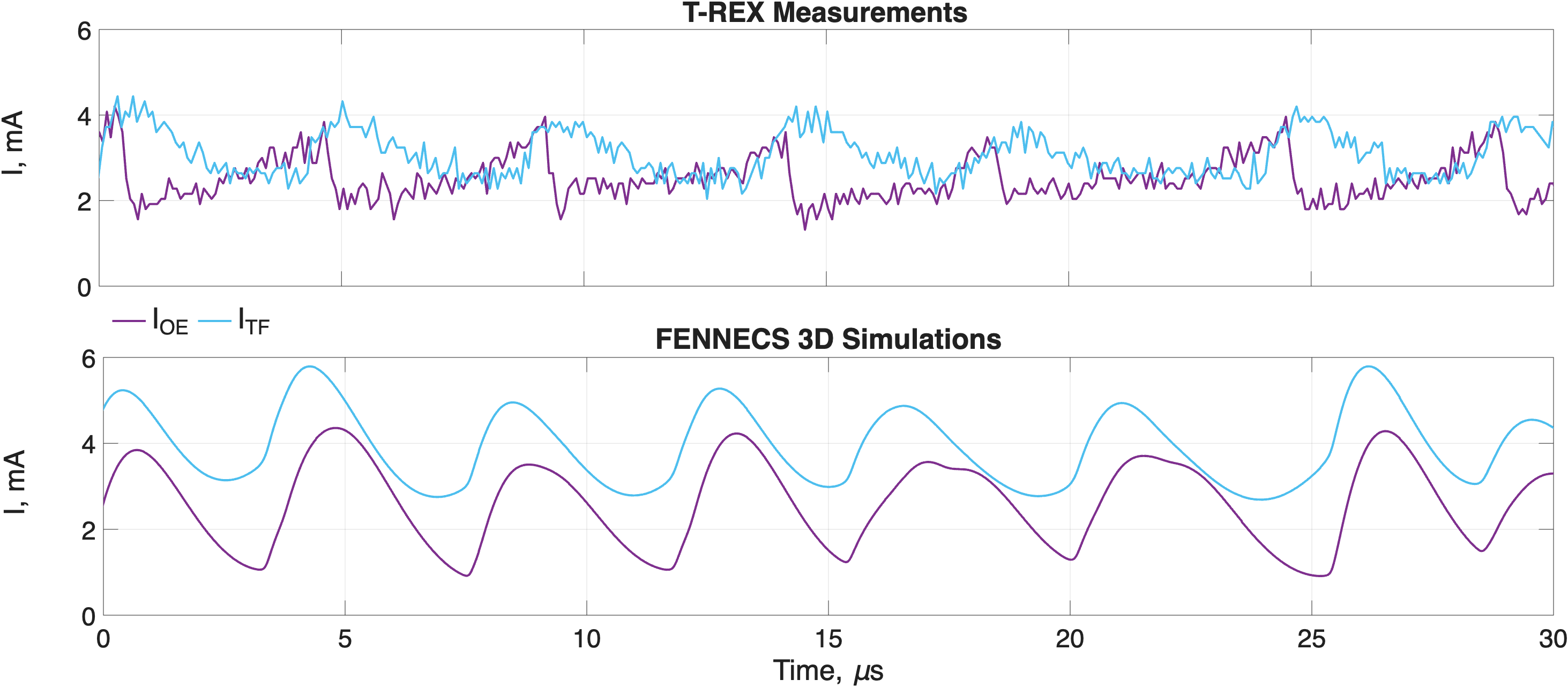}
    \caption{Current amplitude oscillation for both $I_\text{TF}$ and $I_\text{OE}$ vs time for T-REX operating with Ar at $p=\SI{1E-4}{\hecto\pascal}$ for constant $B=\SI{0.10}{\tesla}$ and $\Delta V_\text{bias}=\SI{10}{\kilo\volt}$, vs FENNECS 3D simulation results.}
    \label{fig:currentFENTREX}
\end{figure}

We now show the temporal behavior of the same measurements for the current collected at the top flange $I_\text{TF}$ and at the outer electrode $I_\text{OE}$. The signals are amplified through the same current distribution measurement system characterized by the cutoff frequency $f_c\sim\SI{437}{\kilo\hertz}$ and the accuracy of $\pm\SI{0.05}{\milli\ampere}$. Fig.~\ref{fig:currentFENTREX} shows a comparison between T-REX experimental measurements and FENNECS 3D simulations for $B=\SI{0.10}{\tesla}$. In particular, we observe  current oscillations in the top flange $I_\text{TF}$ and outer electrode $I_\text{OE}$ currents, both experimentally and in simulations, with oscillation amplitudes on the order of $\sim\SI{}{\milli\ampere}$ and at frequencies of $\sim\SI{100}{\kilo\hertz}$. 

This periodicity of the oscillations is identified by the peak oscillation frequency of the signal, or the "burst" frequency $f_\text{burst}$. Physically, the measured burst frequency $f_\text{burst}$ represents the characteristic timescale of electron cloud collapse and reformation driven by the diocotron instability. This process manifests itself as periodic current bursts, followed by a sudden loss of current. We remark that despite the oscillations, the HV amplifier continuously provides the applied bias $\Delta V_\text{bias}$.

\begin{figure}[h]
    \centering
    \includegraphics[width=.6\linewidth]{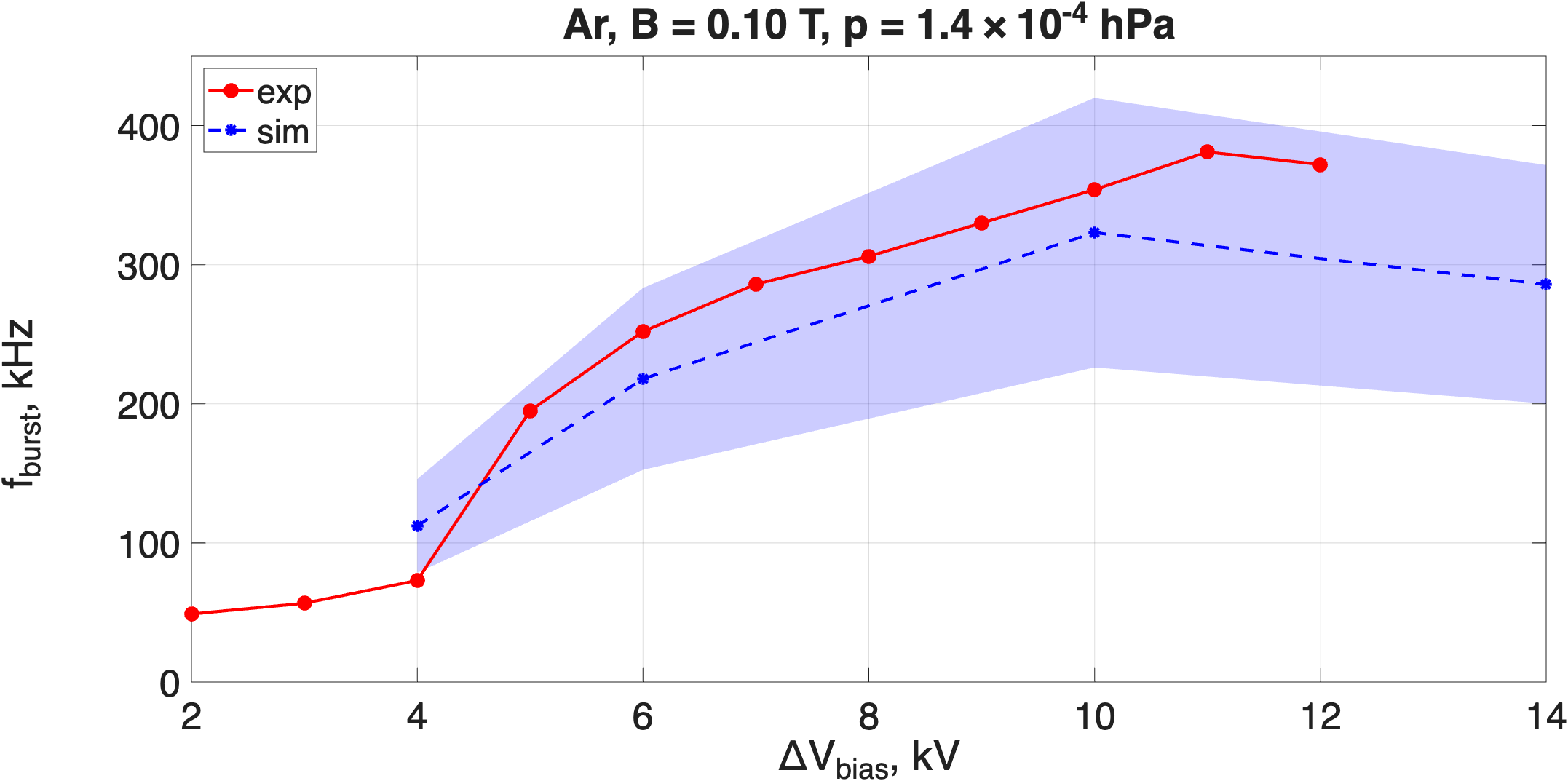}
    \caption{Measured burst frequency $f_\text{burst}$ (derived from the FFT of the current signals) increasing for both $I_\text{TF}$ and $I_\text{OE}$ vs $\Delta V_\text{bias}$ for T-REX operating with argon at $p=\SI{1E-4}{\hecto\pascal}$ for constant $B=\SI{0.10}{\tesla}$, continuous red line, vs FENNECS 3D simulation results with a dashed blue line and shaded area for pressure uncertainty.}
    \label{fig:frequency}
\end{figure}

\begin{figure}[h]
    \centering
    \includegraphics[width=.8\linewidth]{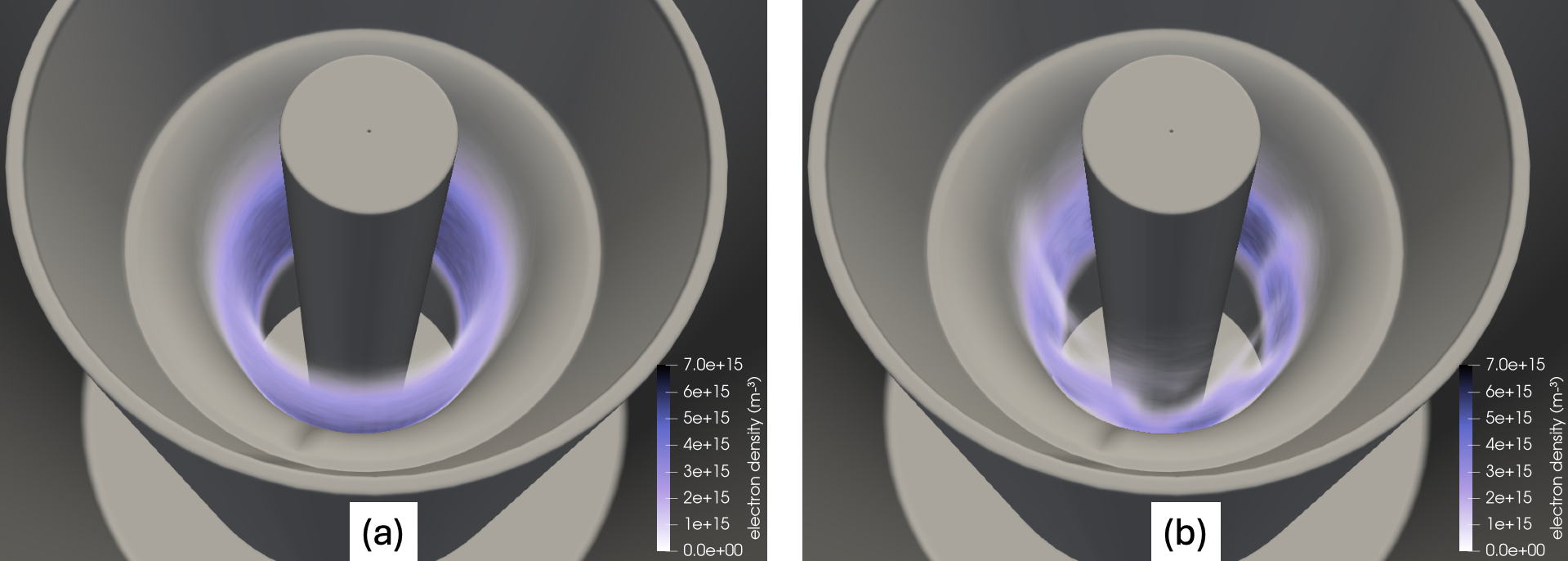}
    \caption{FENNECS 3D Simulation of T-REX: representation in 3D of the electron density before (a) and during (b) a burst due the diocotron instability. Simulated conditions are T-REX with argon at $p=\SI{1E-4}{\hecto\pascal}$, $B=\SI{0.10}{\tesla}$, and $\Delta V_\text{bias}=\SI{10}{\kilo\volt}$. Purple color scale represents the electron density $n_e$. The physical electrodes of T-REX are presented in light grey color.}
    \label{fig:burstnoburst}
\end{figure}

As shown in Fig.~\ref{fig:frequency}, the burst frequency $f_\text{burst}$ for both $I_\text{OE}$ and $I_\text{TF}$ increases with the applied bias $\Delta V_\text{bias}$. This is caused by the larger $\vec{E}\times\vec{B}$ drift velocity at larger applied bias $\Delta V_\text{bias}$, which leads to a greater ionization frequency and, consequently, a higher burst frequency $f_\text{burst}$. Furthermore, a larger applied bias $\Delta V_\text{bias}$ deepens the potential well, in which more charge is then stored. This trend is also corroborated by FENNECS 3D simulations and demonstrates that the code is also able to accurately represent this temporal-behavior. A visualization of FENNECS 3D simulations of T-REX showing the temporal evolution of the electron density before and during a burst is shown in Fig.~\ref{fig:burstnoburst} (a) and (b). This initial agreement provides a strong motivation to investigate such phenomena with more detailed experiments.

\begin{figure}[h]
    \centering
    \includegraphics[width=.85\linewidth]{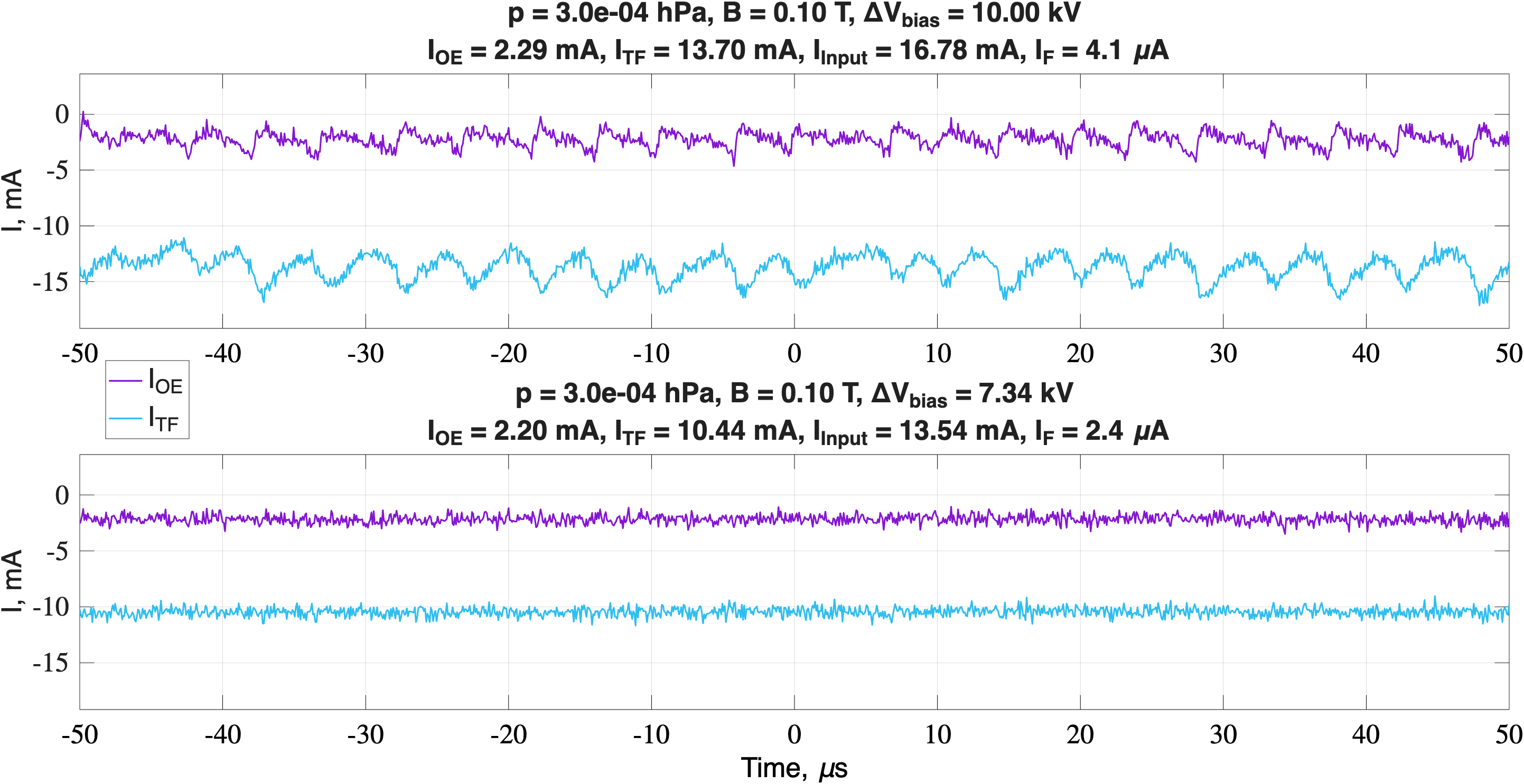}
    \caption{Time-resolved current trace for $I_\text{OE}$ and $I_\text{TF}$ for $\Delta V_\text{bias}=\SI{7.34}{\kilo\volt}$ and $\SI{10}{\kilo\volt}$ for T-REX operating with Ar at for a constant $B=\SI{0.10}{\tesla}$. Bursty mode at the top, quiet mode at the bottom. The two figures highlights the difference in the oscillation amplitude of $\sim2-\SI{4}{\milli\ampere}$ for which a period of \SI{5}{\micro\second} is present for the $\SI{10}{\kilo\volt}$ case, while this is absent for the $\SI{7.34}{\kilo\volt}$ case. Currents are hereby shown negative as the applied bias is also negative.}
    \label{fig:quietmode}
\end{figure}

Furthermore, we also observed conditions for which the amplitudes of oscillations suddenly drop to a minimum. This "quiet-state" occurs only within certain ranges of the applied bias $\Delta V_\text{bias}$, see Fig.~\ref{fig:quietmode}, and does not follow a discernible pattern. Multiple ranges of applied bias $\Delta V_\text{bias}$ present this phenomenon, the only exception is for $B=\SI{0.31}{\tesla}$ where the oscillations remain present regardless of the other experimental parameters. To ensure the results were independent of the experimental facility, we performed several diagnostic tests, including swapping the HV amplifier, varying the HV cable length, and optimizing the grounding and cabling; no such dependencies were observed. FENNECS 3D simulations can also reproduce similar "quiet-states" depending on the initial conditions and the exact value of parameters, but no concrete explanation for these occurrences has been obtained yet. Visually, and also via simulation, it is observed that the electron cloud results more confined radially in the quiet condition compared to the non-quiet one, in which the electron cloud is seen closer to the outer electrode. This phenomenon will further be examined in future work. 

\begin{figure*}[h]
    \centering    
    \includegraphics[width=.9\linewidth]{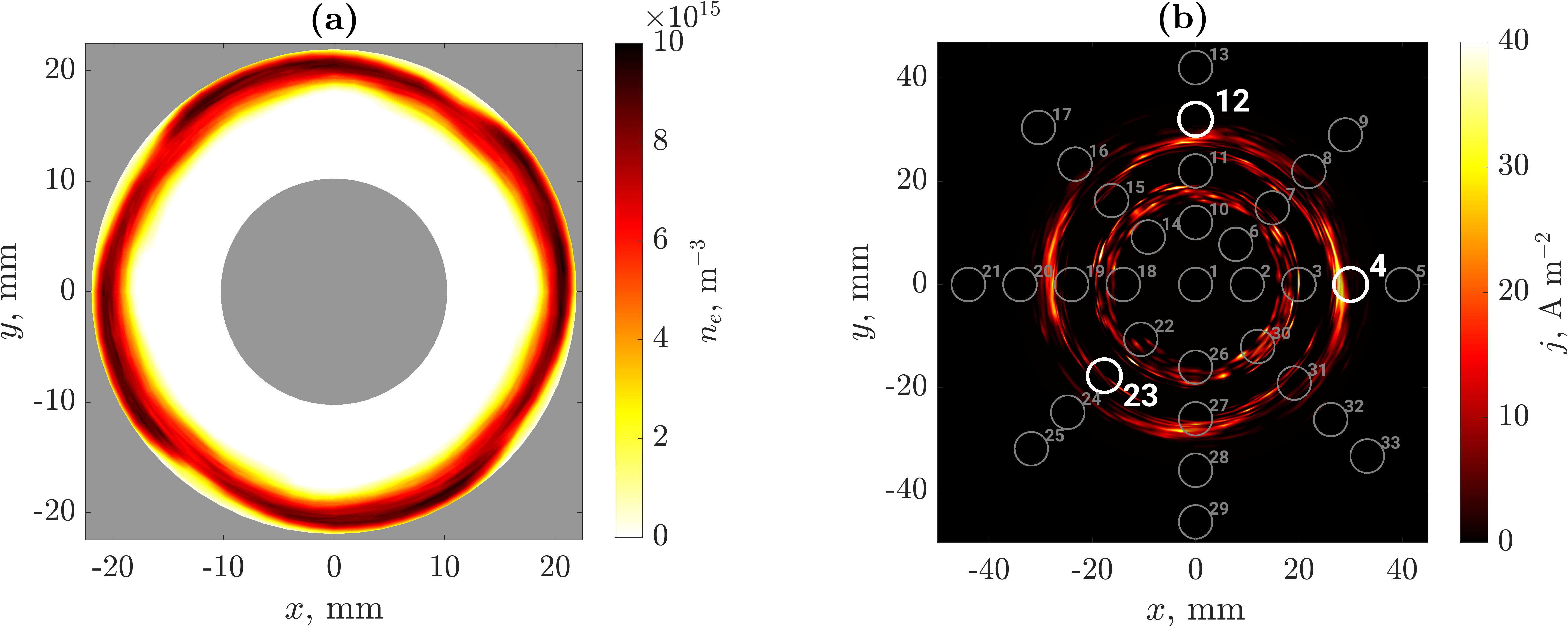}
    \caption{FENNECS~3D simulation of T-REX during a burst, for $B=\SI{0.10}{\tesla}$, $\Delta V_\text{bias}=\SI{10}{\kilo\volt}$ and $p=\SI{1E-4}{\hecto\pascal}$ using argon. The snapshots are taken at $t=\SI{75}{\nano\second}$ during the burst, with the time origin defined in Fig.~\ref{fig:probes_signals_sim}. (a)~Represents the electron density $n_e$ in the midplane of the cloud, showing an $m=4$ azimuthal structure resulting from the nonlinear development of the diocotron instability. The T-REX electrodes are depicted in grey. (b)~Shows the simulated current density $j$ collected on the top flange at the same instant and exhibiting the same $m=4$ azimuthal pattern as the cloud. The grey circles indicate the position of the 33 probes of the current probe array. The fast probes~4,~12 and~23 are highlighted.}
    \label{fig:probes}
\end{figure*}

\begin{figure*}[h]
    \begin{minipage}{0.44\textwidth}
        \includegraphics[width=0.95\linewidth]{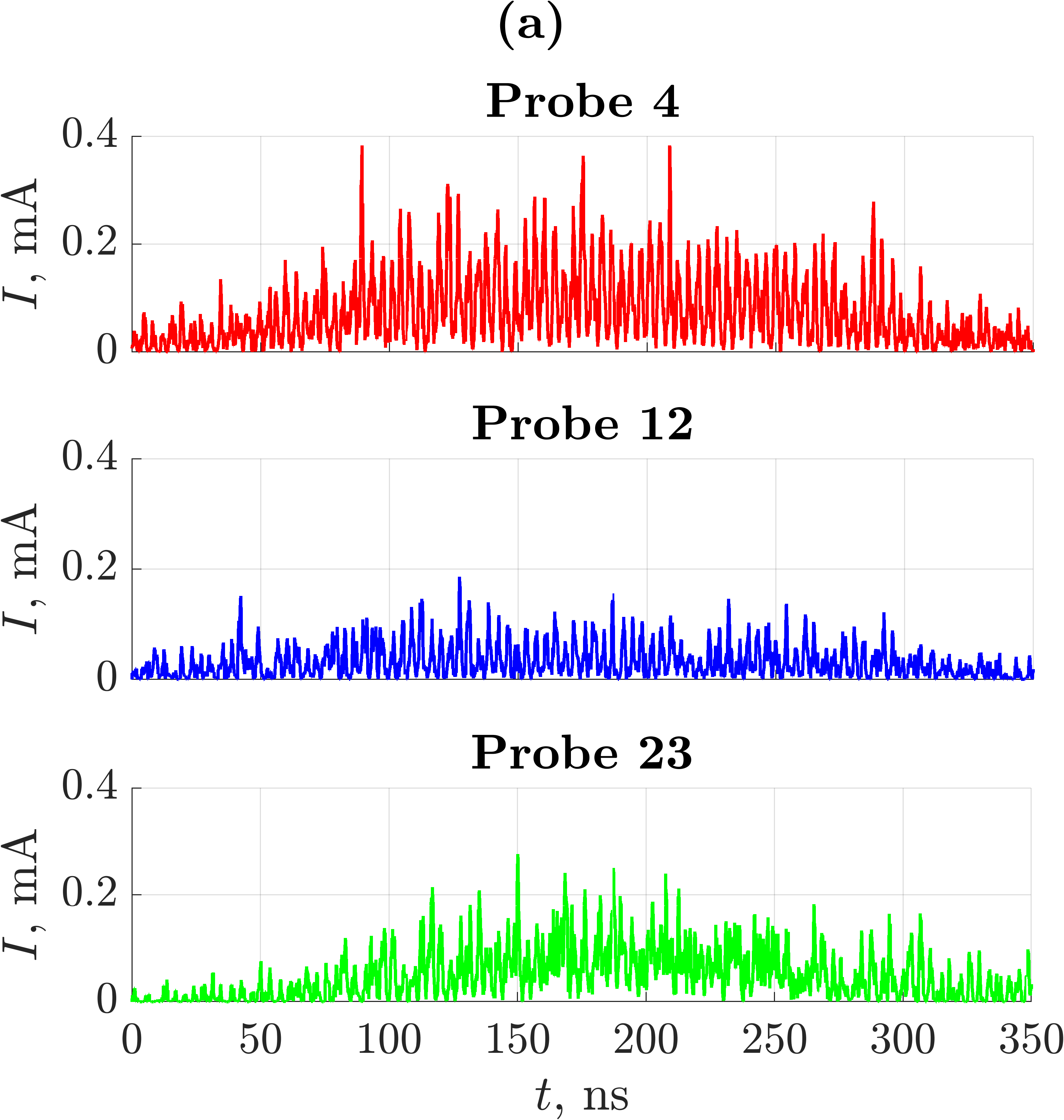}
    \end{minipage}
    \begin{minipage}{0.5\textwidth}
        \includegraphics[width=0.95\linewidth]{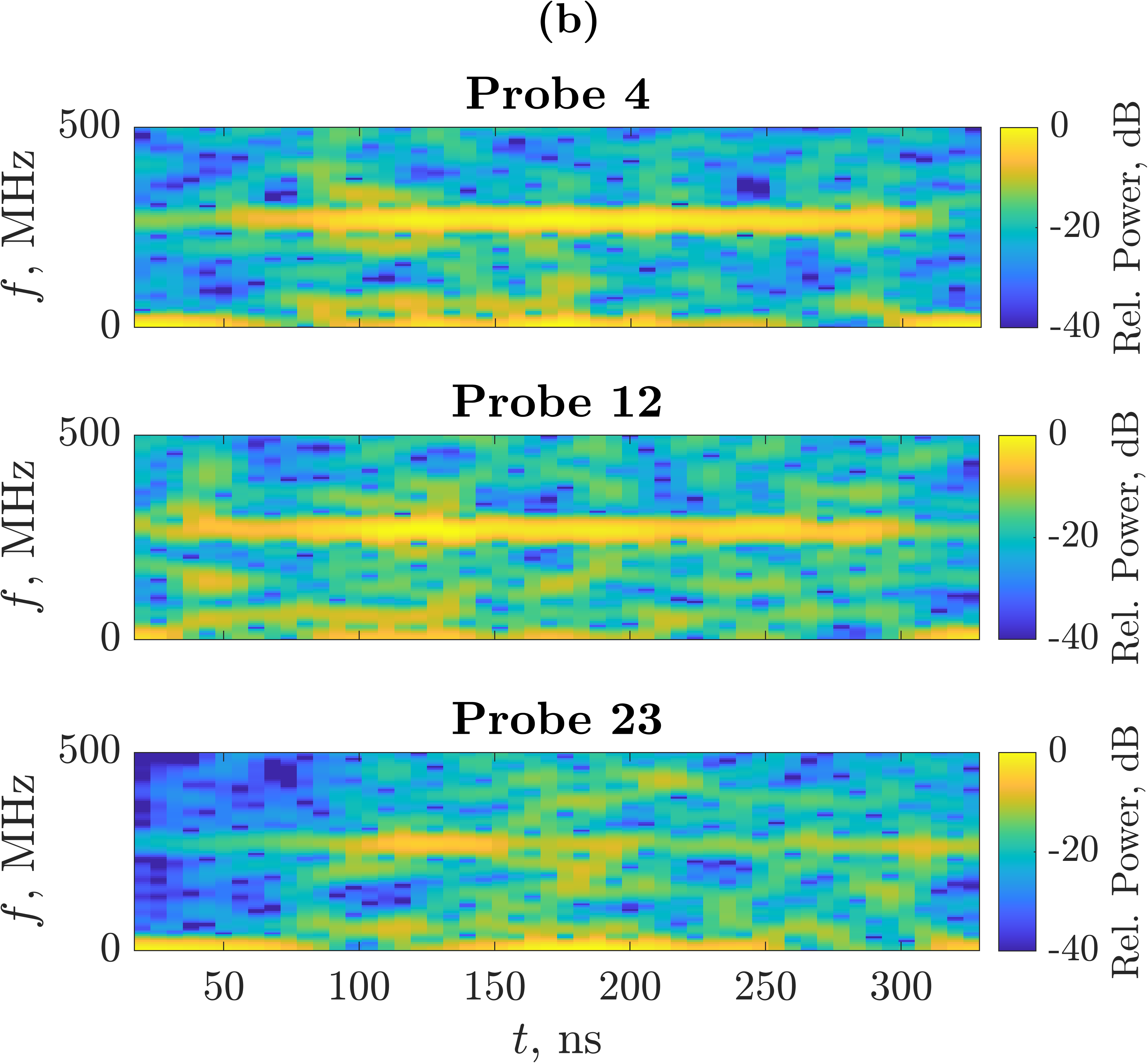}
    \end{minipage}
    \caption{Synthetic diagnostic of the fast probes from FENNECS~3D for the same burst considered in Fig.~\ref{fig:probes} ($B=\SI{0.10}{\tesla}$, $\Delta V_\text{bias}=\SI{10}{\kilo\volt}$, $p=\SI{1E-4}{\hecto\pascal}$, using argon). (a)~Shows the time traces of the synthetic probe currents on probes~4,~12 and~23. (b)~Shows the spectrograms of the synthetic probe signals. The color scale represents the relative spectral power in \SI{}{\decibel}.}
    \label{fig:probes_signals_sim}
\end{figure*}

The primary objective of the current probe array is to detect rotating structures within the electron cloud, specifically those driven by the diocotron instability. Based on synthetic diagnostics implemented in FENNECS~3D, three current probes have been selected to be used with the three available fast amplifiers, namely probes~4,~12 and~23, see Fig.~\ref{fig:probes}. The choice has been made to maximize the collected current and to provide angular separations suitable for the detection of, at least, the $m=\pm1$ mode. The remaining probes are, for the time being, connected directly to the ground without being used. Probes~12 and 4 are separated by an azimuthal angle $\Delta \theta_{12-4} = \pi /2 $, while probes~12 and 23 are separated by $\Delta \theta_{23-12}=3\pi/4$. With this configuration, the smallest angular separation between two probes, $\Delta\theta=\pi/2$ (between probes~4 and~12), allows azimuthal modes up to $m=3$ to be identified through the phase shift $\Delta\phi=m\,\Delta\theta$, which remains below $2\pi$ for $m\leq3$. 

To illustrate the physical mechanism that the probe array is designed to capture, we display in Fig.~\ref{fig:probes} what happens during a current burst of the FENNECS~3D simulation presented in Fig.~\ref{fig:currentFENTREX} performed for $B=\SI{0.10}{\tesla}$, $\Delta V_\text{bias}=\SI{10}{\kilo\volt}$, $p=\SI{1E-4}{\hecto\pascal}$ using argon. Fig.~\ref{fig:probes}~(a) shows the electron density in the midplane of the cloud during the burst, in which a clear $m=4$ azimuthal structure is visible. The nonlinear development of this diocotron mode expels electrons from the potential well while preserving its azimuthal pattern, so that the same $m=4$ structure is imprinted on the current density collected at the top flange, as shown in Fig.~\ref{fig:probes}~(b). Two ring-like features are visible on the top flange current density: the inner ring originates from IIEE at the central electrode, while the outer ring corresponds to the electrons escaping the cloud and discharging axially during the burst, i.e. to the periodic loss of confinement driven by the diocotron instability, which is the target of the present analysis.

As the mode rotates azimuthally with the $\vec{E}\times\vec{B}$ drift, it generates oscillating signals on the probes at a frequency corresponding to the diocotron rotation frequency of the mode. To allow a direct comparison with the experimental measurements, a synthetic diagnostic is implemented as a post-processing step of the FENNECS~3D simulations. At each output time step, the current collected by a given probe is computed by integrating the simulated current density $j$ at the top flange over the surface of the probe, thereby reproducing the quantity measured experimentally. The resulting signals are shown in Fig.~\ref{fig:probes_signals_sim}~(a) for the same burst as Fig.~\ref{fig:probes}. The probes collect a significant current only during the burst, when the electrons are expelled axially toward the top flange. Due to its more inward radial position, probe~23 collects a signal with a different temporal structure compared to probes~4 and~12. The corresponding spectrograms of the synthetic probe signals, see Fig.~\ref{fig:probes_signals_sim}~(b), exhibit a coherent frequency component at $f = 269 \pm \SI{2}{\mega\hertz}$ on all three probes, which persists throughout the duration of the burst. The uncertainty on $f$ corresponds to the FFT bin width. This frequency is consistent with the diocotron rotation frequency of the $m=4$ mode, equal to four times the fundamental $\vec{E}\times\vec{B}$ rotation frequency $f_{\mathrm{sim}, m=1}=\SI{67}{\mega\hertz}$, obtained directly from the cloud drift velocity in the FENNECS~3D simulation during the burst.

The mode number can be retrieved from the relative time delay
$\Delta t$ between the probes using
\begin{equation}
    m= \frac{2\pi f \Delta t}{\Delta\theta},
    \label{eq:mode_estimation}
\end{equation}
with $f$ the dominant frequency of the signal and $\Delta\theta$ the angular separation between the probes. The time delay is estimated by cross-correlation, which operates in the time domain by computing the degree of similarity between the two signals as a function of a trial time-shift $\Delta t$. The value of $\Delta t$ that maximizes the correlation coefficient $0<R<1$ defines the average time-shift of the perturbation between the two probes. For the case considered here, the synthetic diagnostic yields $\Delta t_{12-4}=4.3\pm\SI{0.1}{\nano\second}$ between probes~4 and~12, where the uncertainty corresponds to the sampling interval ($\delta t = 1/f_s = \SI{0.1}{\nano\second}$, with the sampling frequency $f_s = \SI{10}{\giga\hertz}$). The cross-correlation involving probe~23 yields a lower correlation coefficient and does not allow a reliable estimate of the time delay. With $f= 269 \pm \SI{2}{\mega\hertz}$ and $\Delta\theta_{12-4}=\pi/2$, Eq.~\eqref{eq:mode_estimation} gives $m = 2\pi f \Delta t_{12-4} / \Delta\theta_{12-4} = 4.6\pm0.1$. 

This value is close to, but not exactly equal to, the $m=4$ mode directly visible on the electron density and on the top flange current density in Fig.~\ref{fig:probes}. The deviation is most likely due to the fact that probes~4 and~12 are not located at the same radial position: probe~12 is $\SI{2}{\milli\meter}$ further from the center than probe~4. Since $j$ at the top flange exhibits not only an azimuthal but also a radial structure according to FENNECS~3D, the two probes do not sample the rotating mode at the same radial position, therefore introducing a systematic time-delay offset between the two signals that adds to the purely azimuthal contribution accounted for in Eq.~\eqref{eq:mode_estimation}. The mode number inferred from the cross-correlation is therefore biased. This is confirmed by
repeating the synthetic analysis with the two probes placed at the same
radius. The systematic offset then vanishes and the cross-correlation
yields $m=4.0\pm0.1$, in excellent agreement with the $m=4$ structure
directly observed in Fig.~\ref{fig:probes}. This effect equally applies to the experimental measurements and must be kept in mind when interpreting them. The bias could be removed by adapting the probe array so that the fast probes are located at the same radius. The current spiral arrangement was chosen to maximize the radial coverage of the array - its primary design goal - and therefore does not provide pairs of probes at identical radii.

It should also be emphasized that the value obtained here for the diocotron instability mode number is particular to the set of simulation parameters considered. Across the FENNECS~3D simulations performed over the explored T-REX operating range, different dominant mode numbers are observed from one case to another. They remain, however, of low order ($m\leq6$), and no clear trend with $B$, $\Delta V_\text{bias}$ or $p$ has been identified so far.

Having illustrated by means of the synthetic diagnostic how the diocotron instability generates the probe signals, we now turn to the experimental characterization of these rotating structures in T-REX using the current probe array. We present here the results obtained by performing a scan on the applied voltage $\Delta V_\text{bias}= 6,~8,~10,~\SI{12}{\kilo\volt}$ and magnetic field $B=0.10,~0.17,\SI{0.31}{\tesla}$ for argon at a pressure of $p\sim 1-\SI{3e-4}{\hecto\pascal}$. For each acquisition of experimental results, three bursts are manually selected and the signals from current probes 4,~12,~23 and $I_\text{OE}$ are extracted and processed with a dedicated post-processing routine. The extracted information is in terms of current amplitude and time-behavior. To systematically analyze this broad parameter space without redundant detail, we adopt the multi-stage verification workflow established with the synthetic diagnostics shown earlier on.  
\begin{itemize}
\item Identify the frequencies for the fundamental $m=1$ mode through the FFT of individual probe signals; 
\item Use spectrograms to confirm/detect the presence of higher harmonics;
\item Calculate the harmonics based on the measured fundamental $m=1$ mode and compare them with the theoretical ones;
\item Where the diagnostic allows, measurements are used to confirm whether those frequencies are indeed rotations by applying the conditional sampling and cross-correlation techniques explained earlier on.
\end{itemize}

Experimental results show that the $m=+1$ mode can be identified in every case as shown in Tab.~\ref{tab:ExB} by extracting dominating frequencies from the FFTs. The uncertainty in the measured $f_\text{exp}$ is calculated as the standard error of the mean (SEM), defined as $SEM=\sigma/\sqrt{N}$, where $\sigma$ is the standard deviation, and $N$ is the number of measurements. The value of $f_{\mathrm{th}, m=1}$ is the calculated rotational frequency due to the drift velocity in cylindrical geometry, assuming a uniform electric field in the cloud: 
\begin{equation}
    f_{\mathrm{th}, m=1}= \frac{\Delta V_\text{bias}}{B} \frac{1}{2 \pi r_\text{cloud}^2 \ln(R_2/R_1)},
\end{equation}
where $r_\text{cloud}$ is the radial position of the cloud, $R_2$ is the outer electrode radius, and $R_1$ is the central electrode radius. The uncertainty on $f_{\mathrm{th}, m=1}$ is the standard deviation $\sigma$ based on the uncertainty in the radial location of the cloud of $\pm \SI{1}{\milli\meter}$. Moreover, the uncertainty may be larger because the radius of the outer electrode depends on the axial position of the cloud.%, and because variations in the neutral pressure can locally affect the plasma dynamics. 

A systematic correction to $f_{\mathrm{th}, m=1}$ arises from the contribution of the electron cloud to the electric field. FENNECS simulations show that the self-consistent radial electric field in the cloud region, including both externally imposed and space-charge contributions, exceeds the externally imposed field by approximately $10\%$. This value is obtained by computing the local ratio between the total and externally imposed electric fields and averaging it over the cloud region, and is found to be consistent across all investigated parameter sets ($B$, $\Delta V_\text{bias}$). Although this correction is moderate, it increases the effective rotation frequency and explains the systematically higher experimental values compared to the theoretical estimate based on the external field alone. Therefore, we present in Tab.~\ref{tab:ExB} the results taking into account this expected larger electric field into the calculation of $f_{\mathrm{th}, m=1}$. 

Finally, we evaluate the $Z$-score, defined as the number of $\sigma$ the experimental results is far from the mean of the distribution $Z\text{-score} = |f_\text{exp}-f_\text{th}|/\sqrt{SEM^2+\sigma^2}$, that is for all cases with $|Z\text{-score}|<2$ generally considered as good agreement.
\begin{table}[h]
   \centering
      \caption{Experimental results of T-REX for the $m=1$ base rotation frequency $f_\text{exp}$ compared to the theoretical $\vec{E}\times\vec{B}$ rotation frequency $f_{\mathrm{th}, m=1}$ (with increased electric field by $10\%$ due to the effect of the electron cloud) vs applied bias $\Delta V_\text{bias}$ and magnetic fields $B$. The uncertainty of $f_\text{exp}$ is the standard error of measurement (SEM), while for $f_{\mathrm{th}, m=1}$ is the standard deviation $\sigma$ based on the uncertainty in the radial position of the cloud of $\pm \SI{1}{\milli\meter}$. A $|Z\text{-score}|<2$ is generally considered as good agreement.} 
    \label{tab:ExB}
    \begin{tabular}{cccccccccc}
        \toprule
        & \multicolumn{3}{c}{$B = \SI{0.10}{\tesla}$} & \multicolumn{3}{c}{$B = \SI{0.17}{\tesla}$} & \multicolumn{3}{c}{$B = \SI{0.31}{\tesla}$} \\
        \cmidrule(lr){2-4} \cmidrule(lr){5-7} \cmidrule(lr){8-10}
        $\Delta V_\text{bias}$,~\si{\kilo\volt}~\hspace{5pt} & $f_\text{exp}$,~\si{\mega\hertz} &     $f_{\mathrm{th}, m=1}$,~\si{\mega\hertz} & $Z-$score & $f_\text{exp}$,~\si{\mega\hertz} & $f_{\mathrm{th}, m=1}$,~\si{\mega\hertz} & $Z-$score & $f_\text{exp}$,~\si{\mega\hertz} & $f_{\mathrm{th}, m=1}$,~\si{\mega\hertz} & $Z-$score\\
        \midrule
        $6$  & $31.16 \pm 1.59$ & $33.31 \pm 3.35$ & $0.579$ & $19.85 \pm 0.58$ & $19.59 \pm 1.97$ & $0.124$ & $11.34 \pm 0.21$ & $10.74 \pm 1.08 $ & $0.542$\\
      %  $-7$  &            &                &            &                \\
        $8$  & $41.12 \pm 0.12$ & $44.41 \pm 4.47$ & $0.735$ & $24.63 \pm 0.45$  & $26.12 \pm 2.63$ & $0.560$ & $15.87 \pm 0.99$ & $14.33 \pm 1.44$ & $0.886$ \\
       % $-9$  &            &                &            &                \\
        $10$ & $54.85 \pm 0.15$ & $55.51 \pm 3.46$ & $0.189$ &  $31.94 \pm 0.24$ & $32.65 \pm 2.04$ & $0.350$ & $17.51 \pm 0.65$ & $17.91 \pm 1.12$ & $0.310$\\
        %$-11$ &            &                &            &                \\
        $12$ & $65.68 \pm 0.19$ & $66.61 \pm 4.15$ & $0.233$ & $38.17 \pm 0.47$ &  $39.18 \pm 2.44$ & $0.409$ & $20.96 \pm 0.68$ & $21.49 \pm 1.34$ & $0.351$\\
        \bottomrule
    \end{tabular}
\end{table}

The results shown in Tab.~\ref{tab:ExB} match very well the theory within the sources of uncertainties provided and confirm the rotation of the electron cloud according to the $\vec{E}\times\vec{B}$. The rotation clearly follows the predicted trends, namely a positive correlation with the electric field strength (via the applied bias $\Delta V_\text{bias}$) and an inverse dependency on the magnetic field $B$.

\begin{figure}
    \centering
    \includegraphics[width=0.9\linewidth]{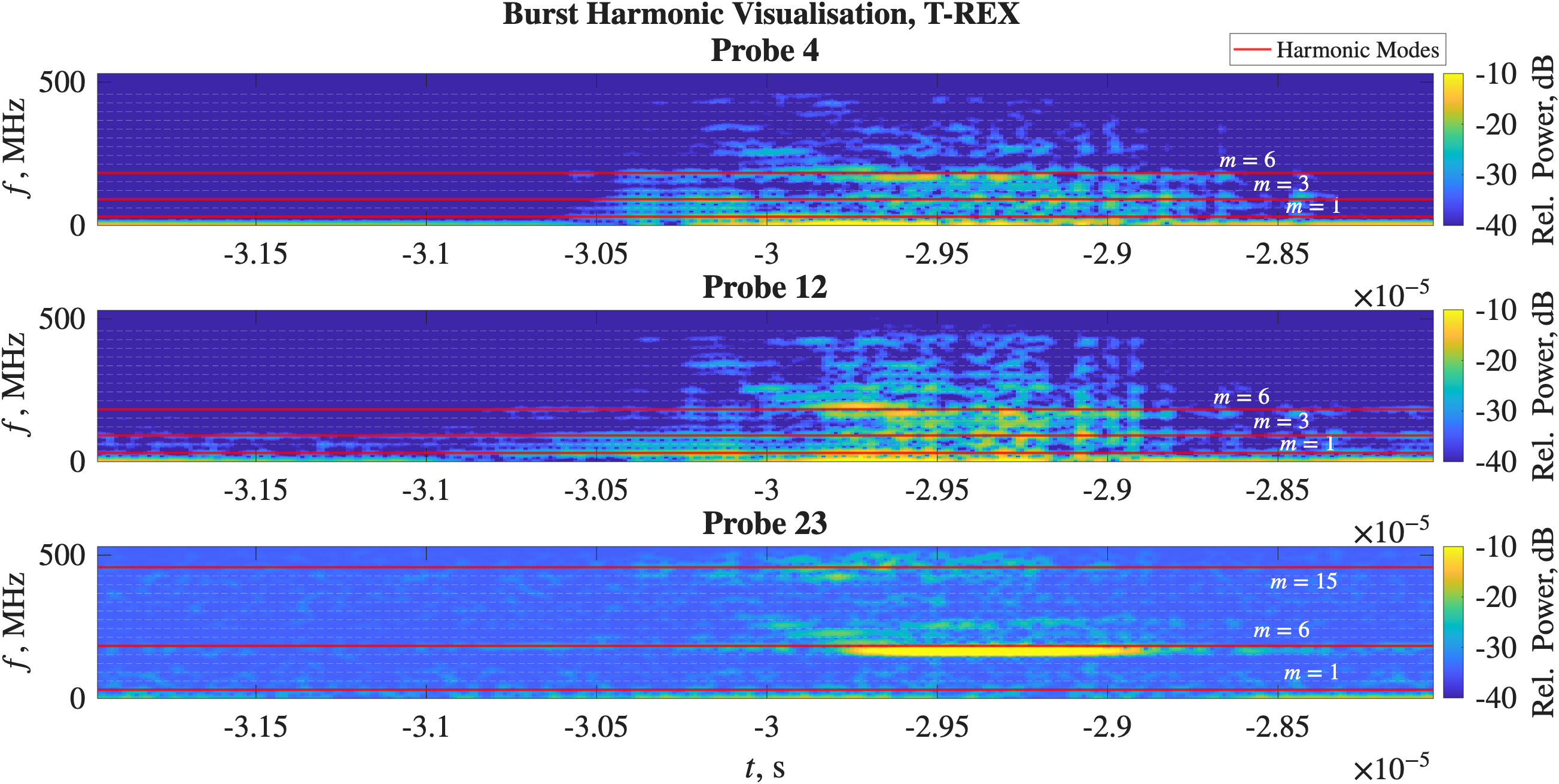}
    \caption{Spectrogram of the signals from Probes~4,~12,~23 for T-REX operating with argon at $\Delta V_\text{bias}=\SI{10}{\kilo\volt}$,~$B=\SI{0.17}{\tesla}$. A burst begins at $t \sim \SI{-3.05E-5}{\second}$ and ends at $t>\SI{-2.85E-5}{\second}$. The red lines with the respective $m$ represent the harmonics $m\times f_{\mathrm{exp}, m=1}$ derived from the measured base frequency of $f_{\mathrm{exp}, m=1}\sim \SI{32}{\mega\hertz}$ as they overlap with the measured ones. The figure includes an extended range on the left of the $x$-axis to provide a baseline, emphasizing the absence of instabilities before the burst is triggered, $t < \SI{-3.05E-5}{\second}$. Colorbar represents the relative power in \SI{}{\decibel}.}
    \label{fig:harmonics}
\end{figure}

Spectrogram analysis highlights that each burst is characterized by a complex multi-frequency spectrum in all three probes. As shown in Fig.~\ref{fig:harmonics}, the harmonic structure of a burst is highlighted and indexed by mode number. From these data, we conclude the following:

\begin{itemize}
    \item The spectra consist of a harmonic series tied to the fundamental $\vec{E} \times \vec{B}$ rotation. The presence of these harmonics allows for the characterization of the rotation independently of cross-probe time-delay measurements;
    \item The application of cross-correlation is constrained by the non-stationary nature of the signals, as distinct frequencies often manifest at different intervals throughout the burst. Such temporal evolution can lead to the identification of "virtual" or spurious modes that do not represent true physical instabilities; consequently, this analysis is limited for the scope of the present study as it needs case-by-case thorough analysis;
    \item The unique frequency signatures detected solely by probe 23 in most cases - the most radially inward probe, see Fig.~\ref{fig:probes} - indicate a radial dependence of the mode structure. This can imply that different rotating structures may be constrained to specific radial envelopes, and also means that it cannot always be used from cross-correlation and conditional sampling together with probes 4 and 12.
\end{itemize}

For further analysis, we compare the extracted peak frequencies from the FFT of the probe signals with the theoretical harmonics as follows: 
\begin{itemize}
    \item Theoretical harmonics are calculated from the theoretical fundamental $\vec{E} \times \vec{B}$ rotation, $f_{\mathrm{th}, m=1}$, as $m\cdot f_{\mathrm{th}, m=1}$, with the associated uncertainty $\sigma_m = m\cdot \sigma_{m=1}$;
    \item Expected harmonics are calculated based on the experimentally measured fundamental $\vec{E} \times \vec{B}$ rotation, $f_{\mathrm{exp}, m=1}$, as $m\cdot f_{\mathrm{exp}, m=1}$ with the corresponding $SEM_m=m \cdot SEM_\text{m=1}$;
    \item The experimentally observed harmonics are plotted and assigned to the closest mode number $m\sim\frac{f_{\mathrm{exp}, m=x}}{f_{\mathrm{exp}, m=1}}$, with the corresponding $SEM_m=m \cdot SEM_\text{m=1}$;
    \item The agreement is evaluated between theoretical predictions and experimental observations across the identified spectrum.
\end{itemize}

\begin{figure}[h]
    \centering
    \includegraphics[width=1\linewidth]{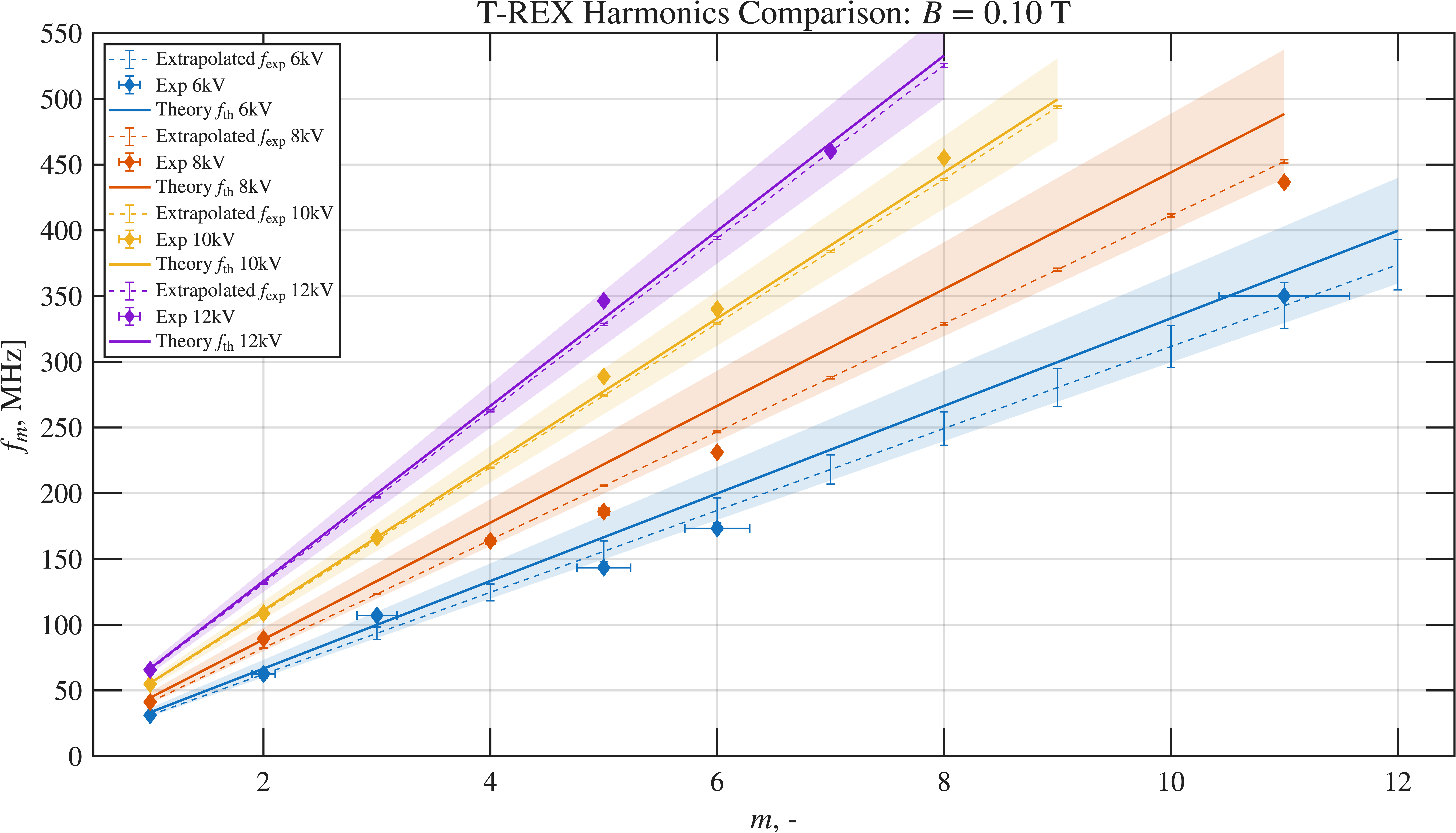}
    \caption{Comparison of experimental and theoretical harmonics for T-REX operating on argon at $B=\SI{0.10}{\tesla}$ for $\Delta V_\text{bias}=6-\SI{12}{\kilo\volt}$. Diamond markers denote experimental measurements; vertical and horizontal error bars represent the SEM and the propagated uncertainty in the harmonic mode $m$, respectively. Dashed lines indicate harmonics extrapolated from the measured fundamental ($f_{\mathrm{exp}, m=1}$), while solid lines represent the theoretical model based on $f_{\mathrm{th}, m=1}$. The shaded regions correspond to the $1\sigma$ theoretical uncertainty.}
    \label{fig:m010}
\end{figure}

\begin{figure}[h]
    \centering
    \includegraphics[width=1\linewidth]{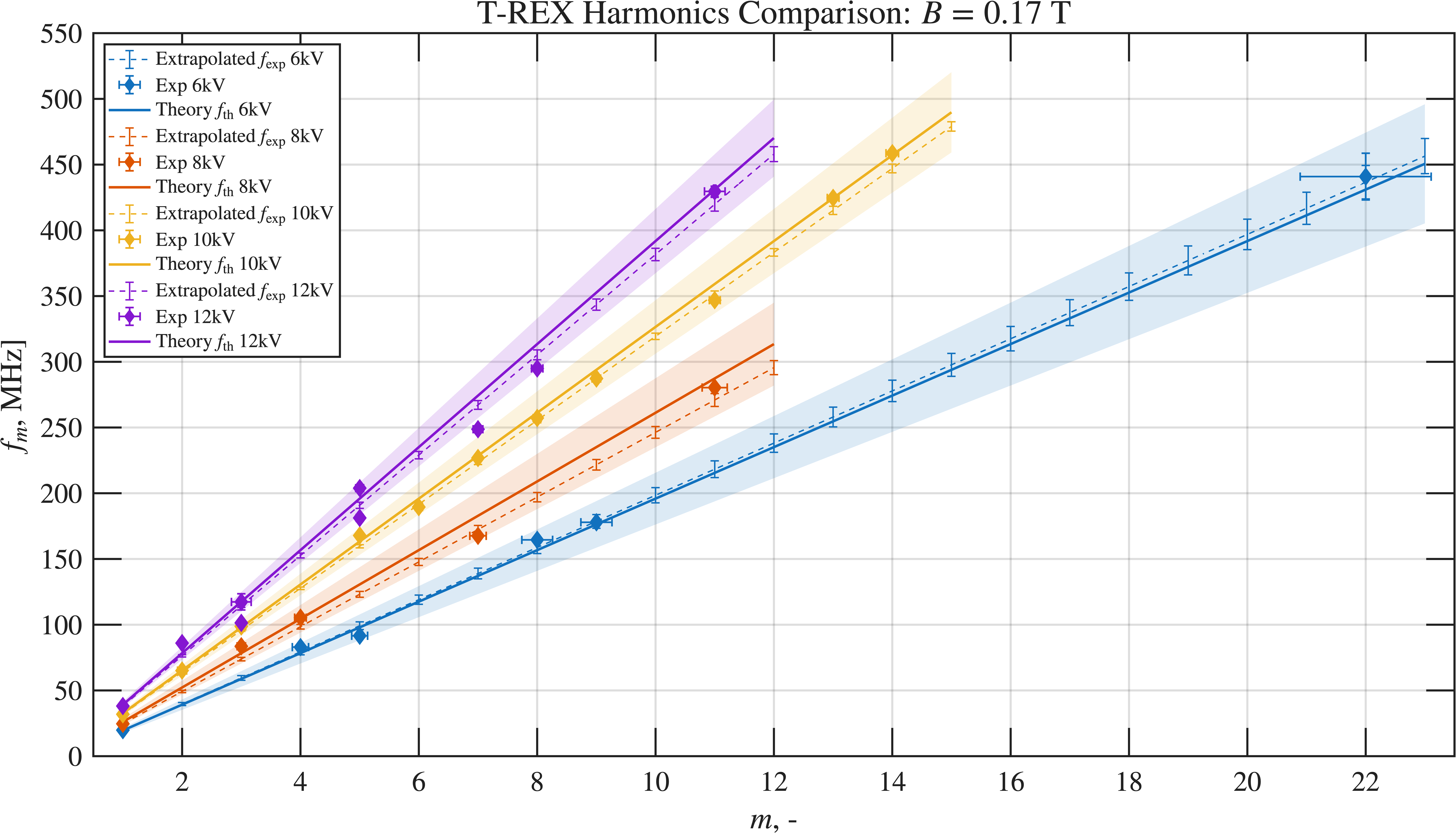}
    \caption{Comparison of experimental and theoretical harmonics for T-REX operating on argon at $B=\SI{0.17}{\tesla}$ for $\Delta V_\text{bias}=6-\SI{12}{\kilo\volt}$. Diamond markers denote experimental measurements; vertical and horizontal error bars represent the SEM and the propagated uncertainty in the harmonic mode $m$, respectively. Dashed lines indicate harmonics extrapolated from the measured fundamental ($f_{\mathrm{exp}, m=1}$), while solid lines represent the theoretical model based on $f_{\mathrm{th}, m=1}$. The shaded regions correspond to the $1\sigma$ theoretical uncertainty.}
    \label{fig:m017}
\end{figure}

\begin{figure}[h]
    \centering
    \includegraphics[width=1\linewidth]{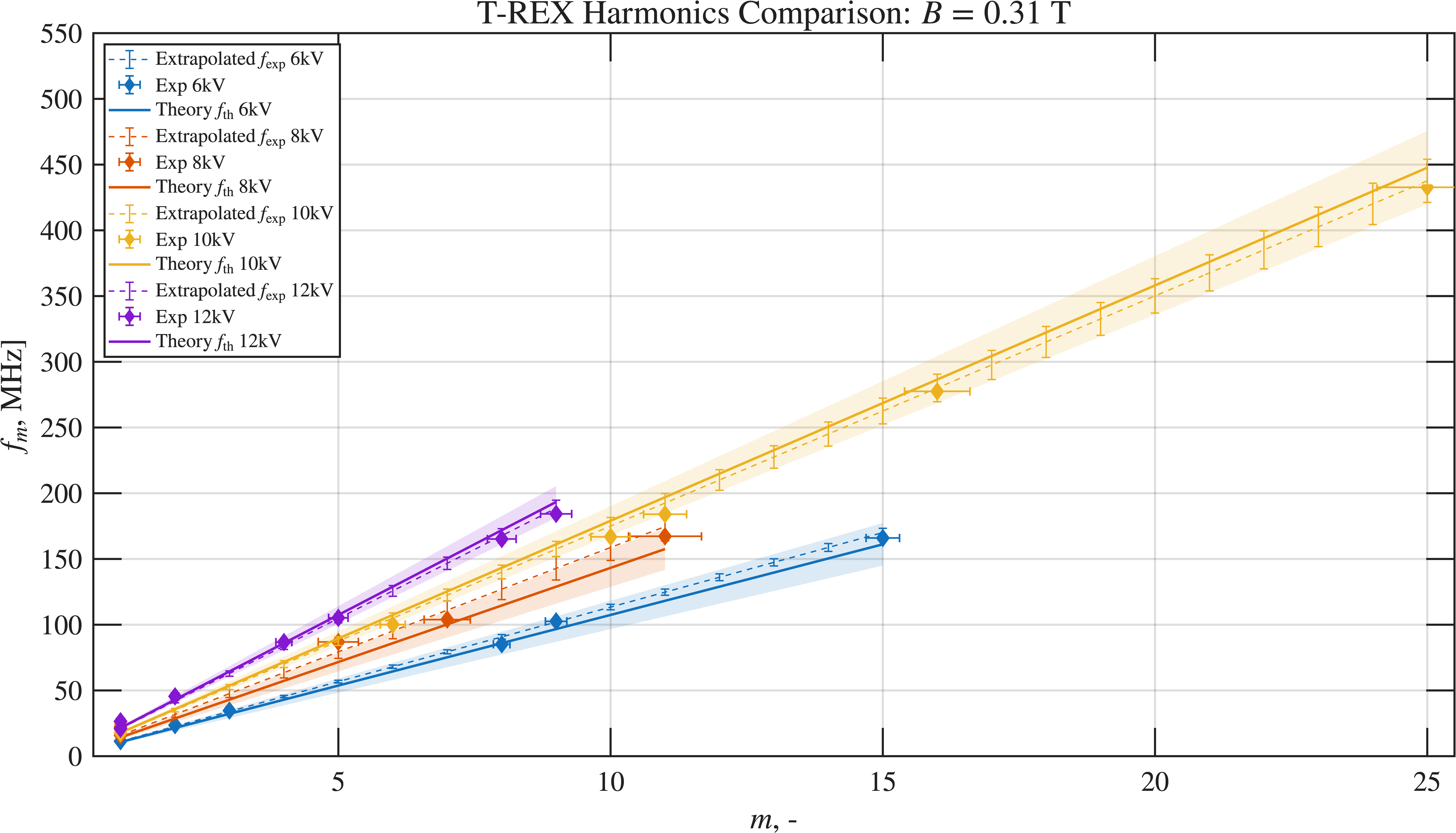}
    \caption{Comparison of experimental and theoretical harmonics for T-REX operating on argon at $B=\SI{0.31}{\tesla}$ for $\Delta V_\text{bias}=6-\SI{12}{\kilo\volt}$. Diamond markers denote experimental measurements; vertical and horizontal error bars represent the SEM and the propagated uncertainty in the harmonic mode $m$, respectively. Dashed lines indicate harmonics extrapolated from the measured fundamental ($f_{\mathrm{exp}, m=1}$), while solid lines represent the theoretical model based on $f_{\mathrm{th}, m=1}$. The shaded regions correspond to the $1\sigma$ theoretical uncertainty.}
    \label{fig:m031}
\end{figure}

The results shown in Fig.~\ref{fig:m010},~\ref{fig:m017},~\ref{fig:m031} demonstrate that the majority of experimental measurements fall within the uncertainty bounds of both the theoretical harmonics, derived from $f_{\mathrm{th}, m=1}$, and the experimental harmonics based on $f_{\mathrm{exp}, m=1}$. Quantitative comparison of estimated slopes confirms a high degree of consistency between theory and experiment, with agreement levels ranging from $89.2\%$ to $98.8\%$. 

These findings verify that the signals detected by the current probes can be harmonics of a single fundamental frequency, providing a more definitive signature of a rotating electron cloud. Consequently, these observations can be identified as a direct trace of the diocotron instability within T-REX. This instability periodically disrupts electron confinement, expelling particles axially toward the top flange. The current probe array, positioned at this boundary, captures these bursts, enabling the real-time characterization of the instability dynamics.

The fundamental theoretical frequency $f_{th}$ is determined by the specific geometry of T-REX and the electron cloud, alongside the primary experimental scaling parameters $B$ and $\Delta V_{\mathrm{bias}}$, the latter increased by $10\%$ to account for the expected electric field amplification due to the electron cloud. However, these results must be evaluated considering the physical assumptions previously discussed. In particular, the lack of direct neutral pressure measurement within the electrode cavity introduces an inherent uncertainty; local pressure variations may influence the burst frequency, leading to over- or underestimation. Furthermore, more detailed analysis shall be performed on the intensity of each peak and, together with the addition of more current probes, it will provide more precise evaluation of the resulting mode $m$. Regarding the statistical representation, the experimental error bars denote the SEM. This choice highlights the precision of the central frequency estimation for the identified modes, though it results in tighter bounds than those defined by the standard deviation $\sigma$.

Preliminary results for the dominant azimuthal mode $m$ are presented for cases validated through both cross-correlation and conditional sampling. This latter statistical technique is based on the identification of a set of reference 'trigger' events, corresponding to the most intense current peaks detected by a reference probe. For each trigger, a temporal window—scaled to three times the inverse of the dominant frequency—is extracted from the signals of all probes and averaged point-wise. This ensemble-averaging process effectively enhances coherent features associated with the rotating physical structure while suppressing uncorrelated background noise. The relative time-delay $\Delta t$ is then determined by comparing the temporal extrema of the resulting conditionally averaged waveforms. By providing a robust representation of the typical temporal structure of the current peaks, this method ensures a reliable estimate of the mean time shift, even in the presence of significant shot-to-shot variability or turbulent fluctuations. For the purpose of this analysis, cases are considered valid only when the cross-correlation coefficient $R>0.55$ and the resulting estimated mode number $m$ is consistent across both methods $m_\text{cross}\approx m_\text{cond}$ for at least two of the three analyzed bursts. This validation specifically utilizes the signals from probes 4 and 12. 

\begin{table}[h]
    \centering
    \begin{tabular}{ccccccccc}
    \toprule
    & \multicolumn{4}{c}{$B = \SI{0.10}{\tesla}$} & \multicolumn{4}{c}{$B = \SI{0.17}{\tesla}$} \\
    \cmidrule(lr){2-5} \cmidrule(lr){6-9}
    $\Delta V_\text{bias}$,~\si{\kilo\volt} & $f_\text{exp}$,~\si{\mega\hertz} & $m_\text{cross}$,- & $m_\text{cond}$,- & $\bar{R}$,- & $f_\text{exp}$,~\si{\mega\hertz} & $m_\text{cross}$,- & $m_\text{cond}$,- & $\bar{R}$,-\\
    \midrule
 6  & $173.21 \pm 0.81$ & $+6.06 \pm 0.03$ & $+6.19 \pm 0.06$ & $0.83$ & NA & NA & NA & NA\\
 8  & $186.10 \pm 1.75$ & $+5.08 \pm 0.10$ & $+5.08 \pm 0.08$ & $0.79$ & $85.31 \pm 0.73$ & $+3.69 \pm 0.01$ & $+3.62 \pm 0.08$ & $0.59$ \\
 10  & $54.85 \pm 0.15$ & $+1.19 \pm 0.02$ & $+1.19 \pm 0.01$ & $0.80$ & $31.94 \pm 0.24$ & $+1.12 \pm 0.11$ & $+1.13 \pm 0.08$ & $0.69$\\
 12  & $65.68 \pm 0.19$ & $+1.16 \pm 0.03$ & $+1.15 \pm 0.05$ & $0.76$ & $166.61 \pm 0.44$ & $+4.88 \pm 0.06$ & $+4.91 \pm 0.01$& $0.68$\\
    \bottomrule
    \end{tabular}
    \caption{Estimation of the azimuthal mode number $m$ and corresponding frequency through cross-correlation ($m_{\text{cross}}$) and conditional sampling ($m_{\text{cond}}$) using probes 4 and 12. Data are shown for a range of $\Delta V_{\text{bias}}$ at $B = \SI{0.10}{\tesla}$ and $B = \SI{0.17}{\tesla}$, while measurements at $B = \SI{0.31}{\tesla}$ yielded no usable signals due to $R<0.55$. For $\Delta V_{\text{bias}} = \SI{8}{\kilo\volt}$ and $\SI{10}{\kilo\volt}$, results are derived from two out of three analyzed bursts where the mode was consistently identified. The parameter $\bar{R}$ denotes the arithmetic mean of the correlation coefficients, and error bars represent the standard deviation $\sigma$ across bursts.}
    \label{tab:m_mode}
\end{table}

These results successfully identify the azimuthal mode number for several of the analyzed cases, providing further evidence that the dominant frequencies detected in each burst correspond to the physical rotational frequencies of the electron cloud. It should be noted that the current error estimate for $m$ is likely an underestimate, as it does not yet account for the uncertainties associated with the probe angular positions $\theta$, the time delay $\Delta t$, and the dominant frequency $f$. Incorporating these factors into the evaluation of $m$ would provide a more reliable assessment of the experimental results. However, for the purpose of this preliminary analysis, rounding the calculated values to the nearest integer is sufficient to identify the mode number. Furthermore, the consistently positive sign obtained for each mode confirms an anticlockwise rotation of the structures. This is in agreement with the expected $\vec{E}\times\vec{B}$ drift direction, dictated by the radial electric field and axial magnetic field configurations specific to the T-REX geometry. While the current setup reliably identifies low-order $m$, the detection and unambiguous characterization of higher-order modes $m>3$ remain constrained by the small number of probes and their spatial distribution. To further confirm and resolve the presence of finer structures within the plasma, a higher density of azimuthal probes would be required to prevent spatial aliasing and improve the resolution of the phase-shift measurements.

\section{\label{sec:conclusions}Conclusions and Outlook}

In this article we have presented an unprecedented set of measurements in the T-REX experiment alongside FENNECS 3D simulations, confirming the central role of the diocotron instability in regulating the disruptive oscillations of trapped electron clouds in MIG-like cavities. 

The results highlight that FENNECS 3D is capable of successfully replicating T-REX experimental current profiles in terms of average amplitude, distribution among the main experiment components, as well as general temporal behavior. Specifically, the code accurately models the burst frequency $f_\text{burst}$ which characterizes the periodic collapse of the electron cloud driven by the diocotron instability. 

Across a range of gases, pressures, and magnetic fields, we observe that currents scale almost linearly with $\Delta V_\text{bias}$. The top flange current $I_\text{TF}$ typically dominates the outer electrode current $I_\text{OE}$, the result of the sum of electrons escaping axially from the electron cloud, as well as electrons that are generated at the central electrode due to ion impact - IIEE. Starting voltage and magnetic field, consistent with previous studies, are at $\Delta V_\text{bias}<\SI{2}{\kilo\volt}$ and $B\sim\SI{0.1}{\tesla}$, leading to a spontaneous formation of the electron cloud. 

The periodic current bursts observed in both $I_\text{TF}$ and $I_\text{OE}$ is a signature of the diocotron instability, which triggers a cloud collapse upon reaching a critical electron density. This is visualized as a sawtooth/sinusoidal current time-trace with a repetition period on the order of hundreds of $\SI{}{\kilo\hertz}$ which tends to increase with the applied $\Delta V_\text{bias}$. Notably, we observed the sudden disappearance of these oscillations, referred to here as "quiet mode", within specific $\Delta V_\text{bias}$ ranges. While this phenomenon also appears in recent simulations and likely points toward an equilibrium state of the electron cloud, it requires more targeted investigation to be fully characterized.

To directly probe the diocotron instability, we designed and implemented a current probe array system at the top of the vacuum chamber to detect rotational structures in the electron cloud. By comparing time-resolved measurements from three high-bandwidth probes - featuring a cutoff frequency $f_c < \SI{1}{\giga\hertz}$ - situated at distinct angular locations, we successfully detected coherent rotational structures. %These measurements identified an $m=+1$ mode rotating at $\SI{30}{\mega\hertz}$, which demonstrates qualitative agreement with FENNECS 3D simulations.

By performing scans of $\Delta V_\text{bias}$ and $B$ we were able to measure and compare the fundamental $\vec{E} \times \vec{B}$ rotation frequency with theoretical values with very good agreement. We further compared measurements of the respective higher harmonics and those also align closely with both theoretical predictions and measured values. 

The combined application of cross-correlation and conditional sampling has allowed for a consistent identification of the dominant azimuthal mode numbers within the T-REX electron cloud in most cases. Specifically, some of the observed frequencies align with the expected $m=+1$ rotational modes measured by single probes, while the consistently positive phase shift confirms an anticlockwise rotation. This direction is physically consistent with the $\vec{E}\times\vec{B}$ drift velocity imposed in T-REX. These findings validate the hypothesis that the estimated $m=1$ frequencies measured on single probes are indeed driven by rotating structures. While the current diagnostic setup is optimized for low-order modes, these results lay the groundwork for future studies aimed at resolving higher-order harmonic structures through increased spatial probe density.

In summary, we report the first direct detection of the diocotron instability within T-REX characterized by rotating structures measured via the current probe array. These experimental observations are in excellent agreement with previous analytical models~\cite{Pierrick} and numerical predictions from both 2D~\cite{guilth} and 3D FENNECS simulations. By leveraging the high-bandwidth current probe array, we have successfully estimated the fundamental rotation frequency of these structures and their dependency on $\Delta V_\text{bias}$ and $B$.
Furthermore, synchronized analysis of these signals has confirmed the existence of well-defined azimuthal symmetries and established the dominant mode number and the rotation direction for some of the cases, providing a definitive link between the observed current bursts and the underlying $\vec{E}\times\vec{B}$ dynamics of the electron cloud. 

Finally, the experimental results from T-REX, supported by FENNECS 3D simulations, provide a comprehensive view of electron cloud dynamics and the onset of the diocotron instability. We have demonstrated that the diagnostic successfully captures complex temporal behaviors, and FENNECS 3D has proven to be a reliable tool for reproducing these non-linear instabilities and basic behaviors. These insights are critical not only for fundamental trapped-electron physics but also for the practical optimization of gyrotron electron guns to meet future high-power requirements. We expect that the measured current bursts are probably responsible for the disruptive events in gyrotrons. For the future, we can apply FENNECS 3D to investigate new geometries and/or mitigation strategies and then utilize the experimental setup of T-REX to test them. 

For the future of T-REX diagnostics, we plan to finalize the current probe array system to enable measurement of the radial electron distribution to further characterize the electron cloud and the diocotron instability, this can also provide the system for an unambiguous $m>3$ detection. The upgrade shall involve optimizing the spatial distribution of the probes and integrating a larger number of them supported by a dedicated amplification and high-speed data acquisition system. Furthermore, we want to further investigate the "quiet-state" by performing more simulation and a detailed experimental campaign including the current probe array system.
In parallel, we are developing an optical spectroscopy system with a radial point of view to measure the electric field within the electron cloud via the Stark effect. These combined experimental efforts will provide the high-resolution data necessary to fully resolve the spatial structure of trapped electron populations.

\begin{acknowledgments}

The authors would like to thank J.~Genoud, M.~Podestà, T.~Goodman, S.~Alberti, and F.~Braunmüller for their insightful contributions during our weekly technical meetings and for their thorough review of the manuscript. We also gratefully acknowledge S.~Antonioni for the development and fabrication of the detection electronics, and M.~Nöel for his expertise in the mechanical design and technical drawings of the T-REX system. A special thanks also to L.~Naux for developing the first design of the current probe array. Finally, we thank the Swiss Plasma Center for fostering the professional environment and providing the facilities necessary to conduct this research.\\

This work was supported by the Swiss National Science Foundation under grant No.~204631.\\

This work has been carried out within the framework of the EUROfusion Consortium, via the Euratom Research and Training Programme (Grant Agreement No.~101052200 - EUROfusion) and funded by the Swiss State Secretariat for Education, Research and Innovation (SERI). Views and opinions expressed are, however, those of the author(s) only and do not necessarily reflect those of the European Union, the European Commission, or SERI. Neither the European Union nor the European Commission nor SERI can be held responsible for them.

\end{acknowledgments}

\nocite{*}
\section*{References}
\bibliography{TREXbib}% Produces the bibliography via BibTeX.

\end{document}